\documentclass[twocolumn]{aastex62}

\sfcode`A=1000 \sfcode`B=1000 \sfcode`C=1000 \sfcode`D=1000
\sfcode`E=1000 \sfcode`F=1000 \sfcode`G=1000 \sfcode`H=1000
\sfcode`I=1000 \sfcode`J=1000 \sfcode`K=1000 \sfcode`L=1000
\sfcode`M=1000 \sfcode`N=1000 \sfcode`O=1000 \sfcode`P=1000
\sfcode`Q=1000 \sfcode`R=1000 \sfcode`S=1000 \sfcode`T=1000
\sfcode`U=1000 \sfcode`V=1000 \sfcode`W=1000 \sfcode`X=1000
\sfcode`Y=1000 \sfcode`Z=1000

\newcommand{\zp}{\ensuremath{z_p}}
\newcommand{\zs}{\ensuremath{z_s}}
\newcommand{\phz}{photo-$z$}
\newcommand{\spz}{spec-$z$}
\newcommand{\0}{\phantom{0}}

\usepackage{booktabs}
\usepackage{graphicx}
\usepackage{natbib}
\usepackage{rotating}

\shorttitle{NIR images and \zp\ in the GOODS-N}
\shortauthors{Hsu et al.}

\begin{document}

\title{Near-Infrared Survey and Photometric Redshifts in the Extended
  GOODS-North field.}

\author[0000-0002-3538-8987]{Li-Ting Hsu}
\affil{Institute of Astronomy \& Astrophysics, Academia Sinica, Taipei, Taiwan}

\author{Lihwai Lin}
\affil{Institute of Astronomy \& Astrophysics, Academia Sinica, Taipei, Taiwan}

\author{Mark Dickinson}
\affil{National Optical Astronomy Observatory, Tucson, AZ 85719, USA}

\author{Haojing Yan}
\affil{Department of Physics and Astronomy, University of Missouri, Columbia, MO 65211, USA}

\author{Hsieh Bau-Ching}
\affil{Institute of Astronomy \& Astrophysics, Academia Sinica, Taipei, Taiwan}

\author{Wei-Hao Wang}
\affil{Institute of Astronomy \& Astrophysics, Academia Sinica, Taipei, Taiwan}

\author{Chien-Hsiu Lee}
\affil{National Optical Astronomy Observatory, Tucson, AZ 85719, USA}

\author{Chi-Hung Yan}
\affil{Institute of Astronomy \& Astrophysics, Academia Sinica, Taipei, Taiwan}

\author{Douglas Scott}
\affil{Department of Physics \& Astronomy, University of British Columbia, 6224 Agricultural Road, Vancouver, BC V6T 1Z1, Canada}

\author[0000-0002-9895-5758]{S.\ P.\ Willner}
\affil{Center for Astrophysics \textbar\ Harvard \& Smithsonian, 60
  Garden St., Cambridge, MA 02138, USA}

\author{Masami Ouchi}
\affil{Institute for Cosmic Ray Research, The University of Tokyo,
  5-1-5 Kashiwanoha, Kashiwa, Chiba 277–8582, Japan} 
\affil{Kavli Institute for the Physics and Mathematics of the
  Universe (Kavli IPMU, WPI), The University of Tokyo, 5-1-5
  Kashiwanoha, Kashiwa, Chiba 277–8583, Japan}

\author[0000-0002-3993-0745]{Matthew L.\ N.\ Ashby}
\affil{Center for Astrophysics \textbar\ Harvard \& Smithsonian, 60
  Garden St., Cambridge, MA 02138, USA}

\author{Yi-Wen Chen}
\affil{National Central University, No.300, Jhongda Rd., Jhongli
  City, Taoyuan County 32001, Taiwan}

\author{Emanuele Daddi}
\affil{IRFU, CEA, Universit\'{e} Paris-Saclay, F-91191
  Gif-sur-Yvette, France}
\affil{Universit´e Paris Diderot, AIM, Sorbonne Paris Cit\'{e}, CEA,
  CNRS, F-91191 Gif-sur-Yvette, France}

\author{David Elbaz}
\affil{IRFU, CEA, Universit\'{e} Paris-Saclay, F-91191
  Gif-sur-Yvette, France}
\affil{Universit´e Paris Diderot, AIM, Sorbonne Paris Cit\'{e}, CEA,
  CNRS, F-91191 Gif-sur-Yvette, France}

\author[0000-0002-0670-0708]{Giovanni G.\ Fazio}
\affil{Center for Astrophysics \textbar\ Harvard \& Smithsonian, 60
  Garden St., Cambridge, MA 02138, USA}

\author{Sebastien Foucaud}
\affil{Shanghai Jiao Tong University, Shanghai, P.\ R.\ China}

\author{Jiasheng Huang}
\affil{National Astronomical observatories of China, Beijing 100020,
  P.\ R.\ China}
\affil{Center for Astrophysics \textbar\ Harvard \& Smithsonian, 60
  Garden St., Cambridge, MA 02138, USA}

\author{David C. Koo}
\affil{University of California Observatories and the Department of
  Astronomy and Astrophysics, University of California, Santa Cruz,
  CA 95064, USA} 

\author{Glenn Morrison}
\affil{LBT Observatory, University of Arizona, 933 N. Cherry
  Ave. Tucson, AZ 85721, USA} 

\author{Frazer Owen}
\affil{National Radio Astronomy Observatory, Socorro, NM 87801, USA}

\author{Maurilio Pannella}
\affil{Faculty of Physics, Ludwig-Maximilians Universit\"at,
  Scheinerstr. 1, D-81679 Munich, Germany}

\author{Alexendra Pope}
\affil{Department of Astronomy, University of Massachusetts, Amherst,
  MA 01002, USA}

\author{Luc Simard}
\affil{National Research Council of Canada, Herzberg Institute of
  Astrophysics, 5071 West Saanich Road, Victoria, BC V9E 2E7, Canada} 

\author{Shiang-Yu Wang}
\affil{Institute of Astronomy \& Astrophysics, Academia Sinica, Taipei, Taiwan}

\begin{abstract}

We present deep $J$ and $H$-band images in the extended Great Observatories Origins Deep Survey-North (GOODS-N) field covering an area of 0.22 $\rm{deg}^{2}$. The observations were taken using WIRCam on the 3.6-m Canada France Hawaii Telescope (CFHT). Together with the reprocessed $K_{\rm s}$-band image, the $5\sigma$ limiting AB magnitudes (in 2\arcsec\ diameter apertures) are 24.7, 24.2, and 24.4 ABmag in the $J$, $H$, and $K_{\rm s}$ bands, respectively. We also release a multi-band photometry and photometric redshift catalog containing 93598 sources. For non-X-ray sources, we obtained a photometric redshift accuracy $\sigma_{\mathrm{NMAD}}=0.036$ with an outlier fraction $\eta = 7.3\%$. For X-ray sources, which are mainly active galactic nuclei (AGNs), we cross-matched our catalog with the updated 2M-CDFN X-ray catalog from \citet{xue16} and found that 658 out of 683 X-ray sources have counterparts. {\it GALEX} UV data are included in the photometric redshift computation for the X-ray sources to give $\sigma_{\mathrm{NMAD}} = 0.040$ with $\eta=10.5\%$. Our approach yields more accurate photometric redshift estimates compared to previous works in this field. In particular, by adopting  AGN-galaxy hybrid templates, our approach delivers photometric redshifts for the X-ray counterparts with fewer outliers compared to the 3D-HST catalog, which fit these sources with galaxy-only templates.\\ 

\end{abstract}


\clearpage
\section{Introduction}

Near-infrared (NIR) observations are essential to understand galaxy
formation and evolution in the distant Universe. NIR data
sample the rest-frame UV to visible light of galaxies beyond the
local Universe and are little affected by dust reddening. These
wavelengths are also a better
tracer of galaxy stellar masses than shorter wavelengths. Moreover,
NIR observations  improve 
photometric redshifts by characterizing the 4000~\AA\
Balmer break  in the spectral energy distribution
(SED) of galaxies at $1<z<4$.

NIR data can be utilized for numerous scientific purposes. For
instance, using the $z_{850}$ dropout technique, we can search for
Lyman break galaxies \citep[LBGs;][] {steidel03} at $z>6.5$ to study
the epoch of reionization \citep[e.g., ][]{bouwens2010,bouwens2015}
by identifying the rest-frame 1216~\AA\ Lyman-$\alpha$ forest
feature. In addition, several NIR color selection techniques can be
used to investigate galaxy properties at high redshift such as
distant red galaxies \citep[DRGs;][]{franx03}, extremely red objects
\citep[EROs;][]{elston88}, dust-obscured galaxies
\citep[DOGs;][]{dey08}, and $BzK$ star-forming and passive galaxies
\citep{daddi04}.

In recent years, many NIR surveys have been carried out with either
ground-based or space telescopes, e.g., the Cosmic Assembly
Near-IR Deep Legacy Survey \citep[CANDELS;][]{grogin11,koekemoer11}
with the WFC3 on the {\it Hubble Space Telescope} ({\it HST}), the
UltraVISTA near-infrared survey using VIRCAM on the VISTA telescope
\citep{McCracken12}, the UKIRT Infrared Deep Sky Survey
\citep[UKIDSS;][]{lawrence07} using WFCAM on the UK Infrared
Telescope (UKIRT), and the Taiwan ECDFS Near-Infrared Survey
\citep[TENIS;][]{hsieh12} using the Wide-field Infrared Camera
\citep[WIRCam;][]{puget04} on the Canada-France-Hawaii Telescope
(CFHT).

The extended Great Observatories Origins Deep Survey-North
\citep[GOODS-N;][]{giavalisco04} is a data-rich field that has been
observed by many observatories in multiple wavebands.
Space observations include X-ray maps with the {\it Chandra X-ray
Observatory} \citep{alexander03}, visible and NIR images
with the {\it Hubble Space Telescope} \citep{giavalisco04},
IRAC 3.6 and 4.5~\micron\ maps in the {\it Spitzer} Extended Deep Survey
\citep[SEDS;][]{ashby13} and deeper
maps in a smaller area with S-CANDELS \citep{ashby2015},
maps at 5.8, 8, 16, and 24~\micron\ with the {\it Spitzer Space Telescope} 
(\citealt{treister2006,teplitz2011}, Dickinson et~al., in prep.),
a far-infrared survey with the {\it
  Herschel Space Observatory} \citep{elbaz11}, and radio observations
with the Very Large Array (VLA) \citep{morrison10}.
While space telescopes are beneficial for
acquiring deeper and higher-resolution images, one of the advantages
of ground-based telescopes is that they can efficiently obtain maps
covering larger areas. Several ground-based NIR observations have
already been carried out in (part of) the GOODS-N region. One of the
first was an eight-band visible to NIR survey  \citep{capak04}.
\citet{wang10a} released a $K_{s}$-band catalog over 
0.25~$\rm{deg}^{2}$ based on observations taken with the CFHT/WIRCam, and
\citet{kajisawa11} published $J$, $H$, and $K_{s}$ data observed
with the MOIRCS instrument on the 8.2~m Subaru telescope in a smaller
area of 103~$\rm{arcmin^{2}}$.

To complement prior data, we have obtained
$J$ and $H$-band images with WIRCam over an 800~$\rm{arcmin^{2}}$
field called the ``Extended GOODS-N Field.''
The primary purpose of this paper is to release the NIR
images, object catalog, and photometry. We also provide estimated photometric
redshifts. Redshifts are essential for most  science
purposes, but measuring spectroscopic redshifts is time-consuming at best.
Moreover, there is a ``redshift desert,'' caused by prominent emission
lines being redshifted out of the visible bands, in which obtaining
spectroscopic redshifts is difficult or impossible. As a result,
only $\sim$4\% of sources in the
extended GOODS-N field have spectroscopic data. Computing  photometric
redshifts in a large survey is a more efficient way to obtain
redshift information. Several photometric redshift catalogs have been
released, e.g., by \citet{rafferty11}, \citet{skelton14}, and
\citet{yang14}. Our work uses the new NIR observations
together with other public data to compute new photometric redshifts
for all of the NIR-detected sources.

X-ray source counterparts are a particularly
important class of objects for which accurate photometric redshifts can help
investigate the correlation between AGN and galaxy evolution in the
early Universe. However,
active galactic nuclei (AGNs) pose special problems for photometric
redshifts because of  their complicated SED
components, and  AGN--galaxy hybrid templates are necessary
\citep{salvato09,salvato11}. \citet{hsu14} took into account varying
ratios of AGN/galaxy contributions and the strong emission lines from
the AGNs to build a set of templates trained on
the sample from the 4~Ms Chandra Deep Field-South (4Ms-CDFS)
X-ray catalog \citep{xue11}. For this paper, we used the most recently updated
2~Ms Chandra Deep Field-North (2Ms-CDFN) X-ray catalog \citep{xue16}
to identify X-ray AGNs and the AGN--galaxy template
library built by \citet{hsu14} to compute photometric redshifts for the AGNs.

The paper is organized as follows. Section~2 describes our new NIR
observations and the collection of the multi-wavelength data used to
compute photometric redshifts. Section~3 shows the photometry
procedure using \textsc{SExtractor} \citep{bertin96} and
\textsc{iraclean} \citep{hsieh12}. We present our photometric
redshifts for non-X-ray and X-ray sources in Section~4. Section~5
presents the column description for the released catalog. Finally we
summarize our results in Section~6.

Throughout this paper, we adopt the AB magnitude system and assume a
flat cosmology with $H_{0}=70$~km~s$^{-1}$~Mpc$^{-1}$,
$\Omega_{\Lambda}=0.7$, and $\Omega _{{M}}=0.3$.

\setcounter{footnote}{0}

\section{Data}\label{datasets}

\subsection{WIRCam Near-IR observation}

\begin{figure}
 \centering
  \includegraphics[trim=0cm 0cm 0cm 5cm, clip=true,width=0.5\textwidth]{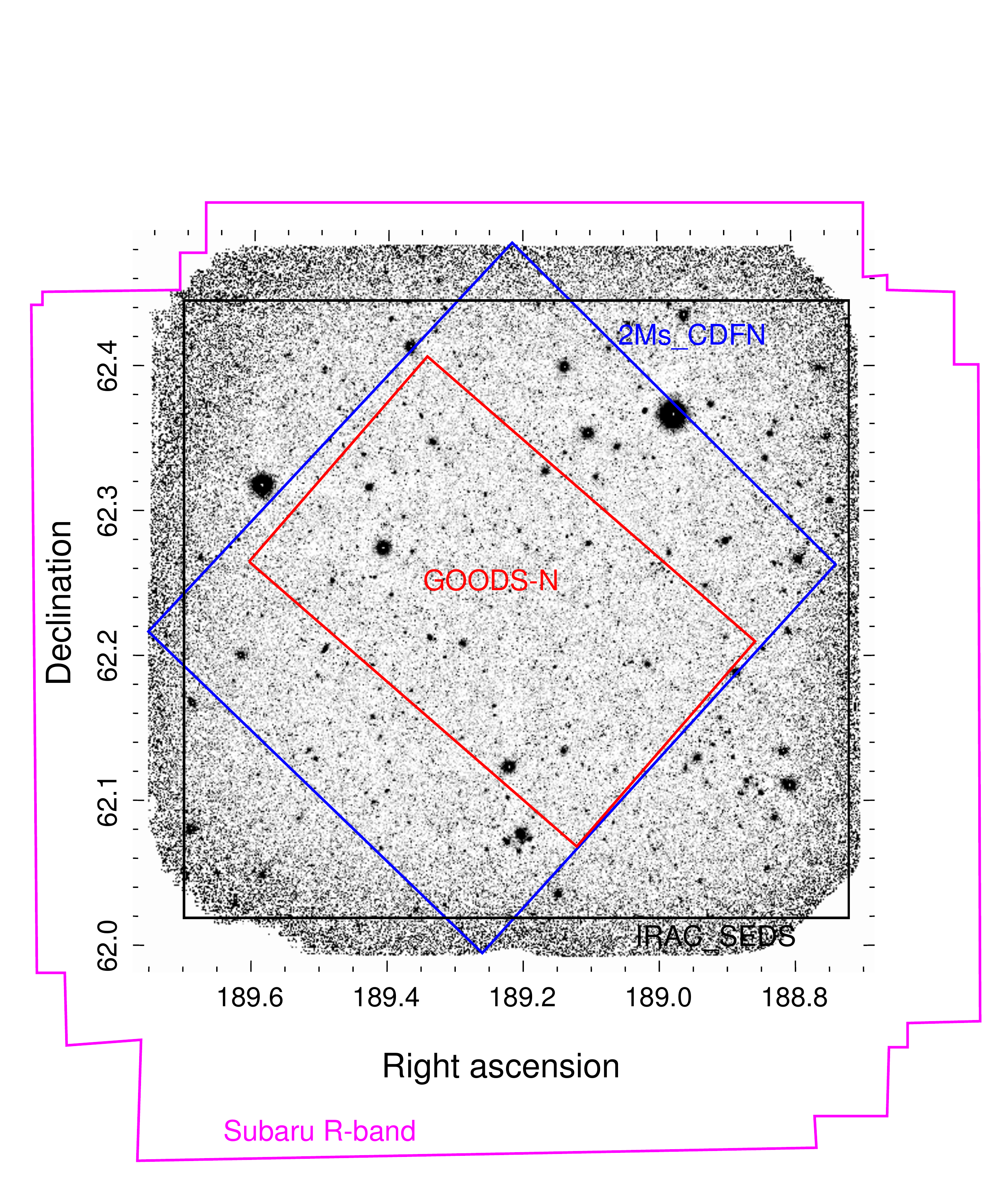}  
  \caption{Observation areas in the extended-GOODS-N field. The
    background image is the CFHT/WIRCam $H$-band image observed in
    this work. The areas from inside to outside are GOODS-N (red
    solid line), 2Ms-CDFN (blue solid line), IRAC/SEDS (black solid
    line), and the Subaru $R$-band image (pink solid line) from
    \citet{capak04}. 
  \label{field} }
\end{figure}

\begin{figure} \centering
  \includegraphics[width=0.45\textwidth]{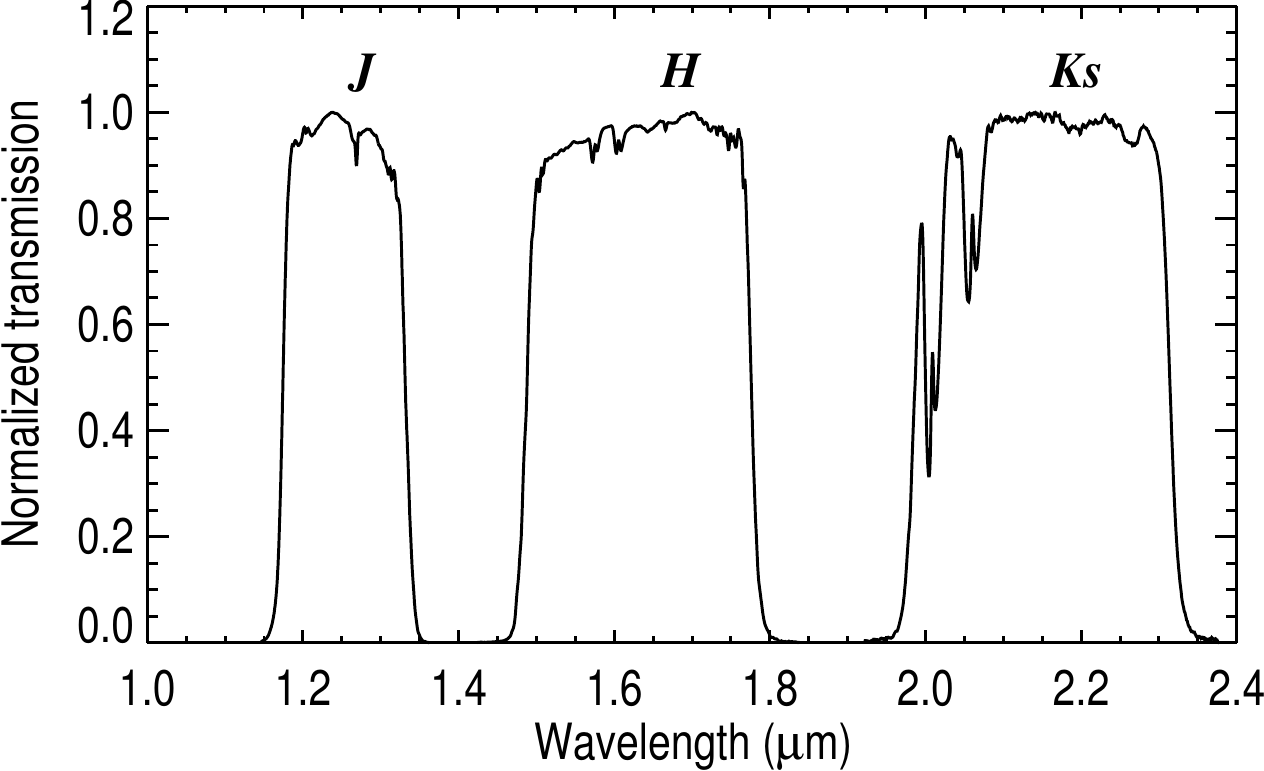}
  \caption{CFHT/WIRCam $J$, $H$, $K_{s}$ filter transmission
curves. The plotted transmissions include the reflectivity of the
primary mirror, transmission of the WIRCam optics, atmospheric
transmission, and detector efficiency. The three curves
are each normalized to a peak value of 1.0.  
   \label{transmission_curve} }
\end{figure}

This work presents new $J$-band and $H$-band observations (PI:
L. Lin) in the extended GOODS-N region (Figure~\ref{field}) together
with $K_{s}$-band data obtained through Hawaiian (PI: L.\ Cowie)
and Canadian (PI: L.\ Simard) programs using WIRCam on the CFHT
\citep{wang10a}. The pixel size of  WIRCam  is
0\farcs3, and the transmission curves of
the three NIR filters are shown in
Figure~\ref{transmission_curve}. All observations were carried out
during 2006--2015 (see Table~\ref{Table:observation_date}), and the
images cover the extended GOODS-N region with area of
28\arcmin$\times$28\arcmin. As  observations  have accumulated during
this long-term project, several studies utilizing these data have
already been carried out 
\citep[e.g.,][]{younger07,keenan10,murphy11,shim11,cassata11,guo12,lee12,penner12,lin12}.

\begin{table*}
\centering
\caption{Observation information \label{Table:observation_date}}
\begin{tabular}{clcccl} 
\tableline
\tableline
Filter  & Semester &Integration time & 5$\sigma$-limit mag
(2\arcsec\ diameter) &Ref. \\ 
\tableline
$J$     &2006A, 2007A, 2009A, 2010A      &47 hrs       &24.7  &
\citet{lin12} \& This work \\ 
$H$    &2011A, 2012A, 2014A, 2015A      &24 hrs       &24.2 &  This work \\
 $K_{s}$  &2006A, 2007A, 2008A, 2009A, 2010A & 52 hrs   &24.4
 &\citet{wang10a} \\ 
\tableline
\end{tabular}
\end{table*}

The observation setups in $J$ and $H$ were similar to those used in
$K_{s}$ \citep{wang10a}. Within each observing block (OB), small
dithering patterns on the order of tens of arcseconds were used to
cover both vertical and horizontal gaps between chips. In each dither
position, two to four subframes were taken with a typical exposure
time of 60~seconds for $J$ and 15~seconds for $H$. In
addition, we offset the centers of dithering by a few arcminutes
after each OB to reduce any systematics associated with different
chips. The typical seeing for
the $J$, $H$, and $K_{s}$ images was between
0\farcs7 and
0\farcs85 (FWHM). The total integration times in $J$, $H$, and
$K_{s}$ are 47, 24, and 52 hours, respectively.

The data were pre-processed using the SIMPLE Imaging and Mosaicking
PipeLinE \citep{wang10b}. The procedure included flat-fielding,
distortion correction, sky subtraction, cross-talk removal, and
photometry calibration against the Two Micron All Sky Survey
\citep[2MASS;][]{skrutskie06} point-source catalog, as described in
detail by \citet{wang10a} and \citet{lin12}. The pre-processed frames
were further astrometrically calibrated against the Sloan Digital Sky
Survey (SDSS) Data Release 6 (DR6) catalog and internal photometry
calibrated using the AstrOmatic software \textsc{Scamp}
\citep{bertin06}. Finally, all the reduced frames were median stacked
using \textsc{Swarp}\footnote{\url {http://www.astromatic.net/}}
\citep{bertin02}. For the final NIR images, the $5\sigma$ limiting
magnitudes (2\arcsec\ diameter aperture) reach $J = 24.7$, $H=
24.2$, and $Ks = 24.4$.

\subsection{ Compilation of UV/visible/IR data}

Taking the advantage of our deep homogeneous NIR images in the wide
field, we assembled publicly released data covering wavelengths from
ultraviolet (UV) to mid-infrared (MIR) to compute photometric
redshifts ({\phz}s). The data we collected are as follows:

\begin{itemize}
\item UV: The far-UV (FUV) and the near-UV (NUV) data were obtained
  from the {\it Galaxy Evolution Explorer} ({\it GALEX}). We matched our
  NIR-detections (i.e., $z$, $J$, $H$, or $K_{{s}}$-detected
  sources, see Section~\ref{sec:detection}) with the {\it GALEX}
  General Release 6/7 (GR6/7). Although the {\it GALEX} image has a
  point-spread function (PSF) of $\sim$5\arcsec, we adopted a search
  radius of 1\arcsec\ to decrease the number of sources with contaminated photometry
  caused by blending. The UV data are used
  only for the X-ray source counterparts (Section~\ref{X_analysis})
  and therefore will not affect sources that are not X-ray counterparts.
  
\item Visible: \citet{capak04} provided $U$, $B$, $V$, $R$, $I$, $z$
  images from the Suprime-Cam\footnote{\url
    {http://www.astro.caltech.edu/~capak/hdf/index.html}} on the
  Subaru telescope with $5\sigma$ limiting magnitudes ranging from 25
  to 27 mag. Additionally, we added $y$-band data taken from the
  Subaru/Suprime-Cam instrument \citep{ouchi09}.
  
\item MIR: We used the IRAC 3.6 and 4.5~\micron\
  data from SEDS
  \citep{ashby13}. IRAC 5.8 and 8.0~\micron\
  data are much shallower than the other two bands, and few
  objects were detected at these wavelengths. 
  For the SED-fitting process, we used
  only $\lambda \le 5$~\micron, as suggested in the \texttt{LePhare}
  manual (page~26), because the stellar light is dominant at these wavelengths.
\end{itemize}

Table \ref{table:photometry} provides detailed information on the
photometric data used to calculate the photometric redshifts. In
all, 14 bands were used for X-ray counterparts and 12 bands for all
other sources.

\begin{table*}
\centering
\caption{Photometric  data\label{table:photometry}}
\begin{tabular}{lllllc} 
\tableline
\tableline
Filter  & $\lambda_{\mathrm{eff}}$ & FWHM& $5\sigma$ Depth (2\arcsec\ aperture)& Instrument/Telescope &Reference\\
&(\AA{})&(\AA{}) &(AB mag) && \\
\tableline
FUV &1539  &228&25.0\tablenotemark{a}      & {\it GALEX}   &General
Release 6/7 \\ 
NUV &2316  &796&25.0\tablenotemark{a}      & {\it GALEX}  &General
Release 6/7 \\  
$U$&3584 &616 &27.03  &KPNO Mayall 4~m/MOSAIC &\citet{capak04} \\
$B$&4374 &1083&27.01   &Subaru/Suprime-Cam&\citet{capak04}\\
$V$&5448 &994&26.40   &Subaru/Suprime-Cam&\citet{capak04}\\
$R$&6509 &1176&26.96 &Subaru/Suprime-Cam&\citet{capak04}\\
$I$&7973&1405&25.85  &Subaru/Suprime-Cam&\citet{capak04}\\
$z$&9195 &1403&25.54 &Subaru/Suprime-Cam&\citet{capak04}\\
$y$&9856  &585               &25.66 &Subaru/Suprime-Cam   &\citet{ouchi09}  \\
$J$&12525 &1568 &24.7   &CFHT/WIRCam& This work\\
$H$ &16335 &2875&24.21   &CFHT/WIRCam& This work\\
$K_{s}$ &21580  &3270  &24.41  &CFHT/WIRCam& \citet{wang10a}\\
3.6~\micron\ &35634  &7444   &25.0\tablenotemark{b}  &
{\it Spitzer}/IRAC & \citet{ashby13}\\  
4.5~\micron\ &45110  &10119  &25.0\tablenotemark{b}  &
{\it Spitzer}/IRAC & \citet{ashby13}\\  
\tableline
\end{tabular}
\tablenotetext{a}{\raggedright
 Values given in Chepter 2 of the GALEX Technical Documentation (\url{http://www.galex.caltech.edu/researcher/techdoc-ch2.html\#2}). }
\tablenotetext{b}{\raggedright
 Values are computed from the rms maps generated by
  \texttt{IRACLEAN}.   
}
\end{table*}

\subsection{X-ray data}\label{xdata} 
X-ray observation is an efficient method to detect active galactic
nuclei (AGN). Studies based on the 4~Ms Chandra Deep Field-South
\citep[4Ms-CDFS;][]{xue11} and 2~Ms Chandra Deep Field-North Survey
\citep[2Ms-CDFN;] []{alexander03,xue16} show that more than 80\%
of X-ray sources are AGNs. As described by
\citet{salvato09,salvato11}, these powerful objects have more complex
SEDs than normal galaxies and therefore require special treatment for
{\phz} estimation. \citet{xue16} provided an updated 2Ms-CDFN
X-ray catalog, which contains 683 X-ray sources including 196
additional X-ray sources compared to the former catalog from
\citet{alexander03}. We used this X-ray catalog to match with our
catalog and identify object to treat as AGNs in the {\phz} estimates.
 
\subsection{Spectroscopic data} 
Spectroscopic redshifts ({\spz}s) were collected from a number of
works, mainly those of \citet{barger08}, \citet{cowie04}, \citet{wang10a},
\citet{cooper2011}, and \citet{xue16}. We matched our NIR detections
(see Section~\ref{sec:detection}) with spectroscopic data within a
maximum separation of 1\arcsec. In total, about 3600 sources
($\sim$4\%) have {\spz} information, and most of them are in the central
GOODS-N region. For more reliable estimates of the {\phz}
quality, we used only 3459 good-quality {\spz}s as indicated by the quality
flag claimed in the literature.

\section{Photometry}\label{sec:photometry}

\begin{table}
\centering
\caption{\textsc{SExtractor} parameters \label{sexpar}}
\begin{tabular}{lc} 
\tableline
\tableline
Parameter &Value \\
\tableline
DETECT\_MINAREA      &2 \\
DETECT\_THRESH       &1.2 \\
ANALYSIS\_THRESH    &1.2 \\
FILTER                           &Y \\
FILTER\_NAME              &gauss\_1.5\_3x3.conv  \\
DEBLEND\_NTHRESH  &64 \\
DEBLEND\_MINCONT   &0.00001 \\ 
CLEAN                           &Y \\
CLEAN\_PARAM            & 0.1 \\
SEEING\_FWHM            &0.8 \\
BACK\_SIZE                   &24 \\
BACK\_FILTERSIZE       &3 \\
BACK\_TYPE                 &AUTO \\
BACKPHOTO\_TYPE    &LOCAL \\
BACKPHOTO\_THICK  &40 \\
WEIGHT\_TYPE            &MAP\_WEIGHT \\
\tableline
\end{tabular}
\end{table}

\subsection{Astrometry and source detection}\label{sec:detection}

\begin{figure}
 \centering
  \includegraphics[width=0.48\textwidth]{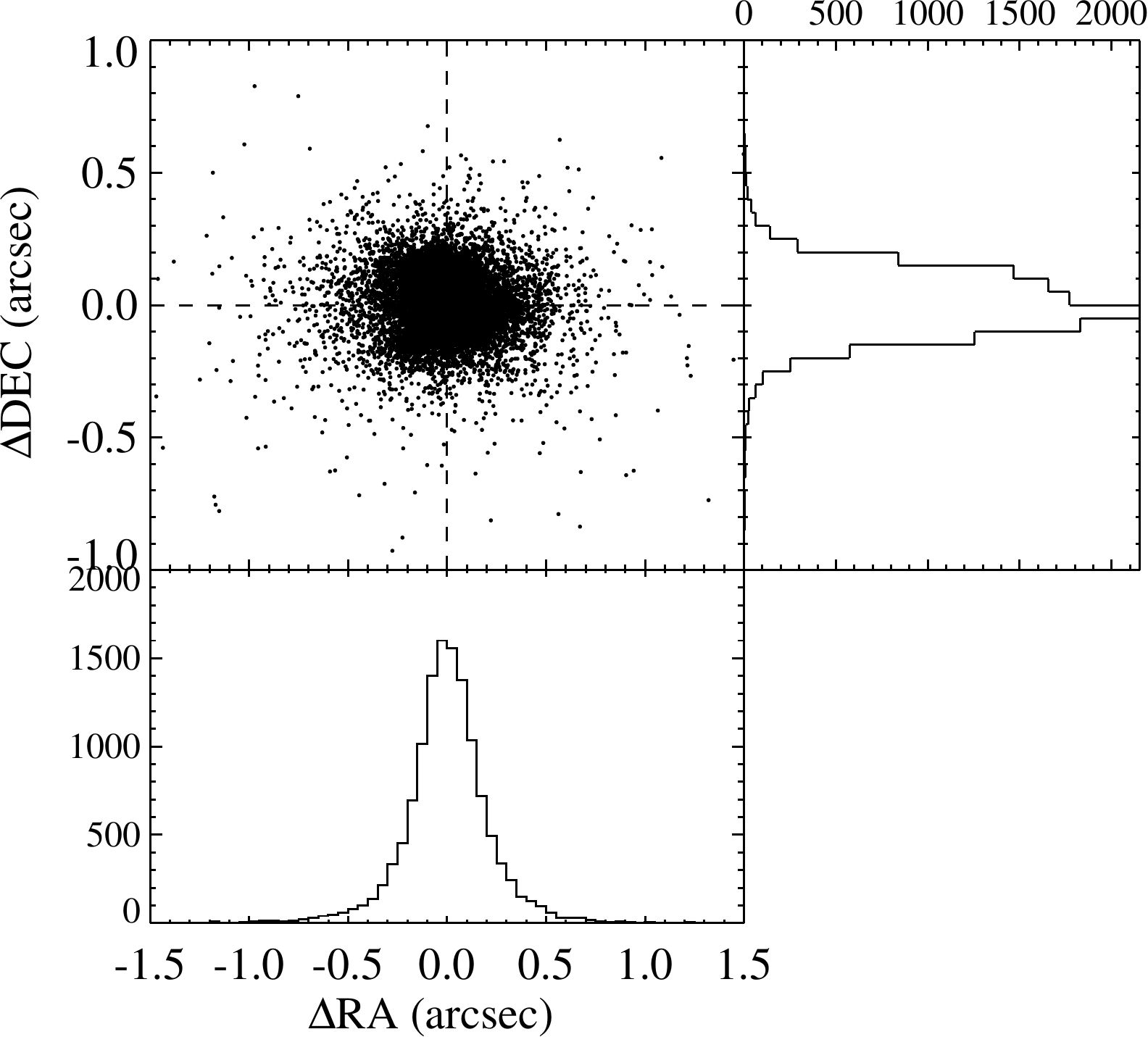}    
   \caption{Astrometric offsets of 12645 sources in common between
     our $zJHK_{{s}}$ catalog and GOODS-N ACS catalog. Upper left
     panel shows offsets for individual sources. Right and bottom
     panels show histograms of offsets in declination and right
     ascension, respectively.   
     \label{fig:delta_radec}} 
\end{figure}

To combine photometric data from different surveys, we have to make
sure that all images are referred to the same astrometric reference
frame. Comparing the coordinates in Capak's visible images with our
NIR images, the median values of systematic offsets are $\sim$0\farcs3--0\farcs4
without any correction. To obtain consistent astrometry,
we aligned all the visible images (i.e., $U$, $B$, $V$, $R$, $I$,
$z$, and $y$) with the WIRCam $K_{{s}}$-band mosaic. After this
calibration, the median values of the systematic offsets between
visible and NIR images range from
0\farcs13 to
0\farcs16. This small offset allows us
to do the source detection and flux extraction for all the images on
the same astrometric reference frame.

Photometry was extracted based on a detection image created
by stacking the non-homogenized NIR (i.e., $J$, $H$, and
$K_{{s}}$) and $z$-band images. These wavelengths are sensitive to UV-luminous
sources at high redshift \citep{laigle16}. The stacked image was
created with the \texttt{CHI2} mode of \textsc{swarp}  \citep{bertin02}.
In this mode, the output image is the
square root of the reduced $\chi^2$ of  pixel values in input
images at a given position.\footnote{\sloppy\texttt{CHI2} mode is set in
  parameter \text{COMBINE\_TYPE}  of
\textsc{Swarp}. See details in Section~6.9.1 of the
\textsc{Swarp} manual: 
{https://www.astromatic.net/pubsvn/software/swarp/trunk/doc/
 \newline swarp.pdf}{.}}
This so-called $\chi^2$ 
image was used as the detection image in the dual-image mode of
\textsc{SExtractor} to define photometric apertures for
each single-band image. In total, 93598 
sources were detected.

For photometry extraction, we used
astrometry aligned to the WIRCam $K_{{s}}$-band
image. However, in the final released catalog, we provide
absolute astrometry aligned with the ACS catalog from
\citet{giavalisco2012}. Figure~\ref{fig:delta_radec} compares
our $zJHK_{{s}}$ detections with the ACS catalog to show the
astrometric offsets. The median values of positional offset in right
ascension (R.A.) and declination (Decl.) are
0\farcs085 and
0\farcs004, respectively.

\subsection{PSF homogenization}

Our 10 ground-based images ($U$, $B$, $V$, $R$, $I$, $z$, $y$,
$J$, $H$, $K_{s}$) have FWHMs of the PSF ranging from
0\farcs8 to~1\farcs3. This will 
give inaccurate flux densities if the same fixed-size aperture is adopted
for all the bands because the ratio of
aperture flux density to total flux density will depend on
PSF size. In order to obtain accurate total flux densities and colors,
which can be used to fit the SEDs and  compute {\phz}s, we need to
take the varying PSFs into account. We adopted the
approach of \citet{capak2007}, degrading the better-seeing images to the
largest PSF (that of the $U$-band image) of 1\farcs3 using a
Gaussian kernel appropriate to each image. To be specific, we used the
PSF characterization (calculating FWHM on images) and equalization
(smoothing with a Gaussian kernel) tools in the
Subaru SuprimeCAM reduction package (\textsc{sdfred}) in
\textsc{iraf}.\footnote{\sloppy{https://www.naoj.org/Observing/DataReduction/mtk/
   \newline subaru\_red/SPCAM/v1.5/sdfred1\_manual\_ver1.5e.html}}

Because the PSF sizes of {\it GALEX} and IRAC  are very large,
to avoid blending, we didn't include UV and IRAC data in
the PSF homogenization. Instead we obtained the UV total flux densities directly
through the {\it GALEX} online query system ``CasJobs''.\footnote{\url
https://galex.stsci.edu/casjobs/}  The IRAC total flux densities
were extracted with the improved \texttt{IRACLEAN}
\citep{hsieh12,laigle16}, which is designed especially for IRAC.

\subsection{Visible/NIR photometry}

\subsubsection{Absolute photometry of NIR images} 
\begin{figure}
 \centering
  \includegraphics[width=0.4\textwidth]{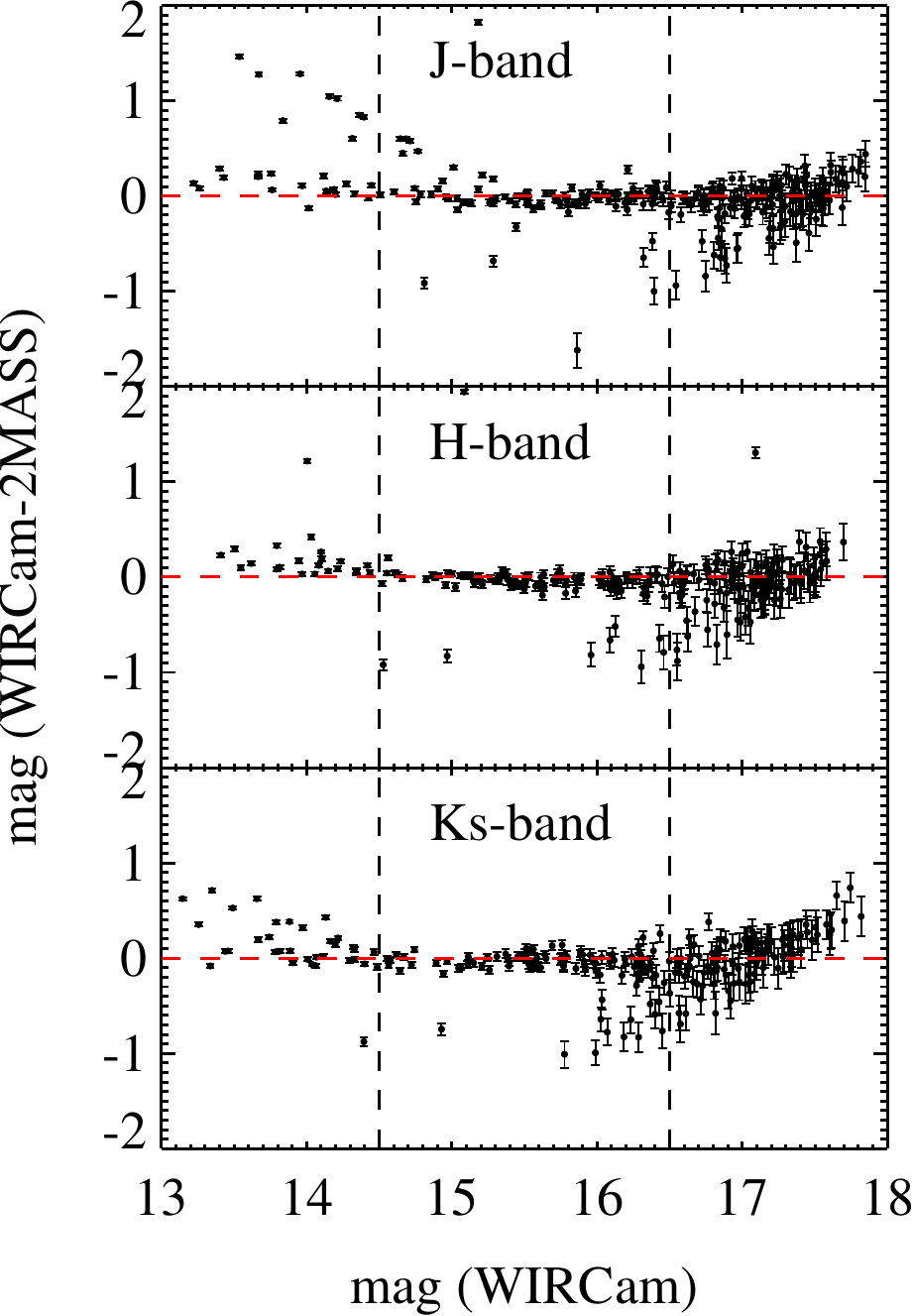}    
   \caption{Magnitude differences between 2MASS and the
calibrated WIRCam AB magnitude in $J$, $H$, $K_{s}$. The
1$\sigma$ combined uncertainties are plotted as vertical error bars.
The horizontal red
dashed-lines indicate magnitude differences of zero. The two vertical black dashed-lines indicate magnitude = 14.5 and 16.5. 
   \label{fig:zjhk_2mass}}
\end{figure}

The absolute
photometry of our WIRCam $J$, $H$, $K_{s}$ images was calibrated
with 2MASS (i.e., The Two Micron All Sky Survey) point sources in
the observed field. Figure~\ref{fig:zjhk_2mass} shows the magnitude differences
between the 2MASS and the calibrated WIRCam magnitudes.
The scatter becomes larger at mag $<14.5$ and $>16.5$. The magnitude differences are mostly constant between 14.5~mag (the brightest WIRCam
sources) and 16.5~mag (where 2MASS $S/N$ deteriorates). We therefore chose the median value in the range $14.5<\rm{mag}<16.5$ to set the calibration for each WIRCam NIR image. In the $J$-band, there are some sources with large scatter at $J<15.5$. Most of these are saturated stars, and their WIRCam photometry is not reliable. However, these sources do not affect our calibrations. 


As mentioned by \citet{wang10a}, the magnitude uncertainties in
Figure~\ref{fig:zjhk_2mass} are dominated by 2MASS. Therefore
our NIR flux calibration is limited by the 2MASS systematic uncertainty.

\subsubsection{$U$ to $K_s$ flux extraction} \label{sec:extr}
As mentioned above, $U$ to $K_s$ photometry was performed with \textsc{SExtractor} in
dual-image mode. We used the $zJHK_{{s}}$ $\chi^2$ image as
the detection image and measured aperture photometry on each
PSF-homogenized image. Fixed-aperture flux densities (i.e.,
\texttt{FLUX\_APER}) have higher signal-to-noise ratio than the
automatic-aperture flux densities (i.e., \texttt{FLUX\_AUTO})
\citep{kron1980}. Therefore, fixed-aperture photometry
gives ``cleaner'' colors for SED fitting and  more
accurate {\phz}s. Tests showed that 2\arcsec\ \texttt{FLUX\_APER}
gave better \phz\ quality (i.e., better accuracy and fewer outliers) than either 
\texttt{FLUX\_AUTO} or 3\arcsec\ \texttt{FLUX\_APER}. 
We therefore used
2\arcsec-aperture photometry for  {\phz} computation. However,
both 2\arcsec\ \texttt{FLUX\_APER} and \texttt{FLUX\_AUTO} are included in the
released catalog.

The \textsc{SExtractor} parameters we used are mainly adopted from
\citet{wang10a} and \citet{hsieh12}, who observed with
the same instrument (i.e., CFHT/WIRCam) as we did.
The parameter \texttt{CLEAN\_PARAM} in particular is critical for source
detection. We examined values of \texttt{CLEAN\_PARAM} from 0.1
to 1.0 and found that the number of detections increased from ${\sim}
90000$ to ${\sim} 130000$. Setting $\texttt{CLEAN\_PARAM} =0.1$  avoids a
large number of false detections as demonstrated by
\citet{yang14}. We also visually inspected random sources
on the images and confirmed that  false detections
are greatly reduced with a lower value of
\texttt{CLEAN\_PARAM}. Table~\ref{sexpar} lists the main
\textsc{SExtractor} parameters  used to obtain the photometry. 

Because {\it GALEX} and IRAC photometry is provided as total flux density, to combine
them with visible and NIR photometry, we need to convert
visible and NIR aperture flux densities to total flux densities for each band using the
equation:
\begin{equation} \label{eq:aper_to_total}  {f_{tot} = f_{\rm aper}
\times {\rm median}\left(\frac{f_{\rm auto}}{f_{\rm aper}}\right) \times 1.06}
\quad .
\end{equation}
First we converted the fixed-aperture flux densities
($f_{\rm aper}$) to automatic-aperture flux densities ($f_{\rm auto}$) by multiplying
the median ratio between \texttt{FLUX\_AUTO} and
\texttt{FLUX\_APER}. \texttt{FLUX\_APER} misses
about 5~to 10\% of the flux density, and
therefore we multiplied by the factor 1.06, the same value adopted by
\citet{yang14}.\footnote{\sloppy The factor is obtained according to Section
  10.4 of the \textsc{SExtractor} manual \newline {
https://www.astromatic.net/pubsvn/software/sextractor/trunk/doc
 \newline /sextractor.pdf}
.}

\subsubsection{Photometric uncertainties}

It is essential to have accurate photometric uncertainties for measuring
accurate {\phz}s. However the uncertainties generated by
\textsc{SExtractor} are usually underestimated because of
correlated noise among pixels. Therefore,
following \citet{bielby2012}, we
computed the rms of flux densities measured in random blank sky
positions with 2\arcsec\ 
apertures. Blank sky positions were identified
from the segmentation map generated by \textsc{SExtractor}.
Then we calculated the ratio between the derived rms in the field and
the mean value of the \textsc{SExtractor} uncertainties of all
objects. These ratios---1.5, 1.72, and 1.78 in $J$, $H$,
and $K_s$ bands, respectively---should be the correction factors. To
be conservative, we applied a correction 
factor of 2 to the \textsc{SExtractor} uncertainties in all the bands
before computing the {\phz}s\footnote{The photo-$z$ quality (indicated by accuracy and outlier fraction, defined in Sec.~4) becomes significantly worse only when the correction factors are greater than 4.}. This
correction is {\em not} included in the photometric errors in the released
catalog, which gives the errors output by \textsc{SExtractor}.

\subsection{IRAC photometry}

IRAC images have PSFs with FWHM around
1\farcs8 (IRAC Instrument
Handbook\footnote{\sloppy
{http://irsa.ipac.caltech.edu/data/SPITZER/docs/irac/
 \newline iracinstrumenthandbook/IRAC\_Instrument\_Handbook.pdf}}),
larger than the visible and NIR bands. Therefore they require
different procedures for deblending objects and determining the
flux densities accurately. Several methods have been developed
for performing IRAC flux density measurements. In this work, we used the
algorithm \texttt{IRACLEAN} \citep{hsieh12}. The method was further
improved \citep{laigle16} to obtain more accurate flux densities for blended
objects with large flux differences and separations less than one
FWHM. Unlike other methods that first deconvolve high-resolution
images with high-resolution PSFs and then convolve with the
low-resolution PSFs, the \texttt{IRACLEAN} approach merely
deconvolves the IRAC low-resolution images with the IRAC PSFs and
adopts source positions and surface brightnesses as priors to derive
the IRAC flux densities. Without requiring identical morphologies for each
object (as other methods do), this method minimizes the effect of
morphological k-correction (see \citealt{hsieh12} and
\citealt{laigle16} for details).

In the \texttt{IRACLEAN} procedure, we used the high-resolution
WIRCam $zJHK_{{s}}$ $\chi^{2}$ image as a prior for the
low-resolution IRAC photometry. The IRAC flux densities were registered to
the associated detections in the stacked-$zJHK_{{s}}$ segmentation
map generated by \textsc{SExtractor}. The uncertainty was
estimated from fluctuations in the local area for each object in the
residual map, taking the non-independence of mosaic pixels into
account.

\subsection{Completeness} 
\begin{figure*} \centering
   \includegraphics[trim=0cm 6cm 0cm 7cm, clip=true,
width=0.9\textwidth]{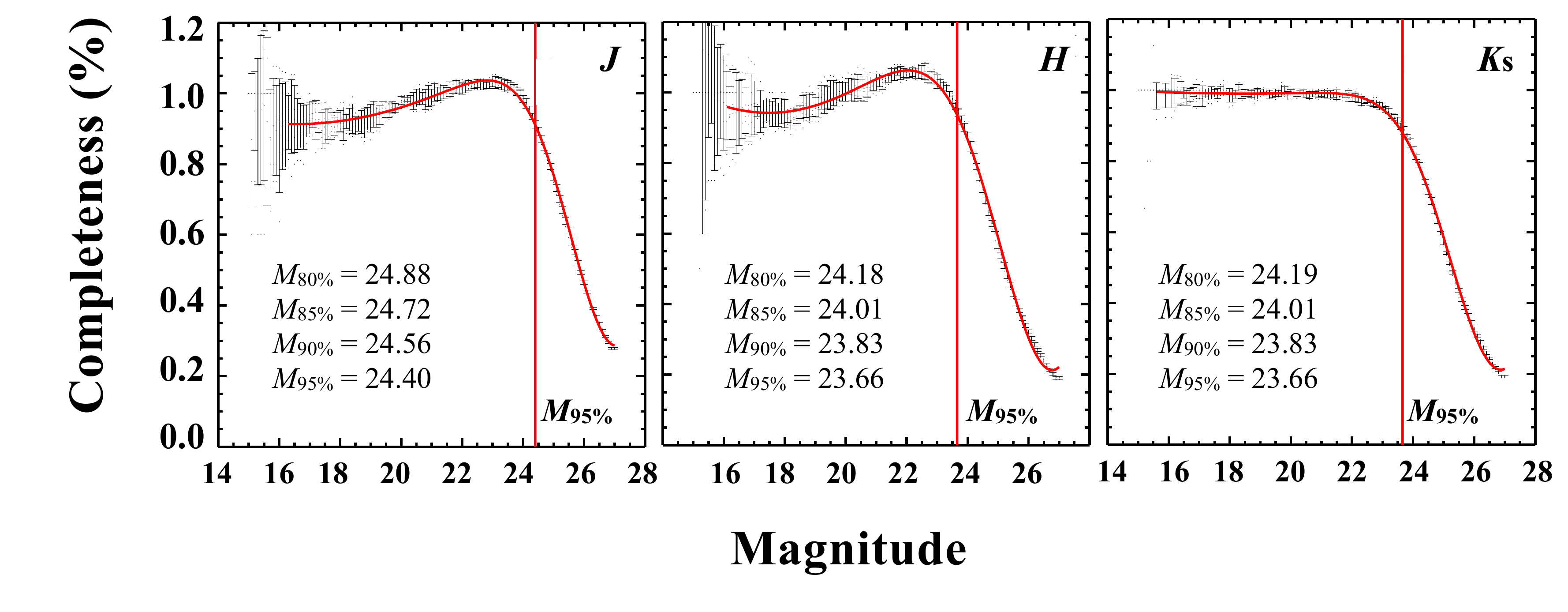}
   \caption{Completeness estimated by simulated stars for the $J$,
$H$, and $K_{s}$ band data (fitted with red solid curves from
left to right panels). The red vertical lines mark the limiting
magnitudes for $95\%$ completeness. Magnitudes for other completeness
values (from $80\%$ to $90\%$ ) are given in each panel.\\
\label{completeness} }
\end{figure*}

We used simulations, performed separately
for each WIRCam filter, to characterize the completeness functions of
the WIRCam observations. Artificial stars of different magnitudes
were generated and added to the reduced WIRCam images. Then we
applied the same source extraction technique to detect and extract
photometry. These artificial stars were recovered or missed
depending on their flux, local noise, crowding, etc. Thus the
completeness of photometry in each filter was evaluated as a function
of magnitude based on the fraction of recovered artificial stars
input in the reduced image
(Figure~\ref{completeness}). The limiting magnitudes corresponding
to 95\% completeness are 24.4, 23.7, and 23.7 in the $J$, $H$ and
$K_{s}$ bands, respectively. 

\begin{figure*} \centering
   \includegraphics[width=0.85\textwidth]{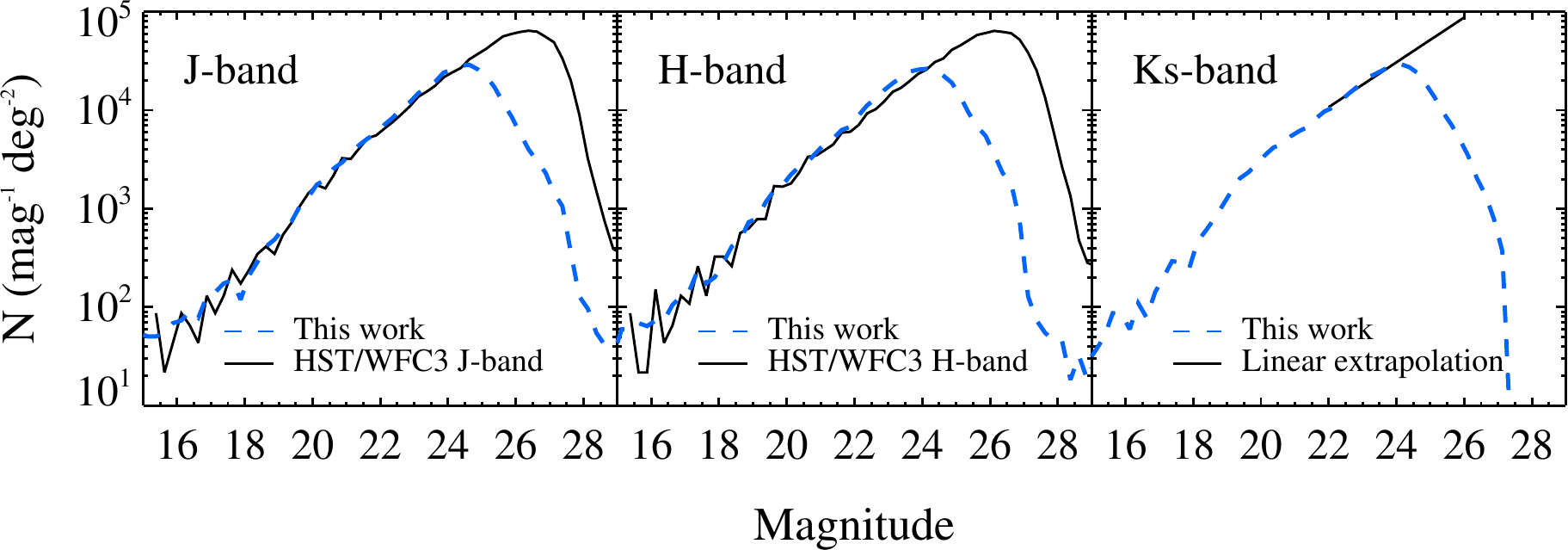}
      \caption{Comparisons of number densities between our work (blue dashed-lines) and the 3D-HST catalog (black solid lines) in $J$ and $H$ bands. In $K_{s}$ band, the black solid line shows the linear curve extrapolated based on the best fit in the magnitude range  from 22 to 24 mag.
      \\
\label{completeness_ext} }
\end{figure*}

In fact, most of our detections are extended sources. To be more realistic, for $J$ and $H$ bands, we compared the number density of our detections with the deeper HST/WFC3 observation from the 3D-HST catalog \citep{skelton14} to deduce the completeness (Figure~\ref{completeness_ext}). We assumed that the 3D-HST catalog is complete at magnitudes less than 26 in $J$ and $H$ bands, and our catalog reaches 99\% completeness in $J$ and $H$ bands at 24.5 and 24.2 mag, respectively. Indeed the number densities of our work at some points are higher than the 3D-HST catalog. This could be due to the different filter bandpasses. For $K_{s}$ bands, however, no public image is deep enough for us to do this measurement in this field. Therefore we extrapolated from the $K_{s}$-band number density to estimate 99\% completeness at 23.9 mag.

\section{Photometric redshifts}\label{galphz}

We computed {\phz}s using the publicly available code
\texttt{LePhare} \citep{arnouts99,ilbert06}, which is based on
$\chi^{2}$ minimization for obtaining the best-fit template. We
divided the source catalog into non-X-ray-detected and X-ray-detected
subsamples and applied an appropriate template library for each
subsample separately. For non-X-ray-detected sources, we fitted with
pure galaxy templates, whereas for the X-ray-detected sources, we
fitted with AGN-galaxy hybrid templates. In addition, we fitted both
subsamples with stellar SED templates to identify stars. In the
following {\phz} quality analysis, we excluded not only
spectroscopically-confirmed ({\spz} = 0) stars but also
SED-classified stars defined by $\chi_{\rm star}^{2} < \chi_{\rm
best}^{2}$. In the whole spectroscopic sample, there are 240
SED-classified stars, and 196 ($\sim$82\%) of them are
spectroscopically confirmed. 

\begin{figure}
 \centering
  \includegraphics[width=0.4\textwidth]{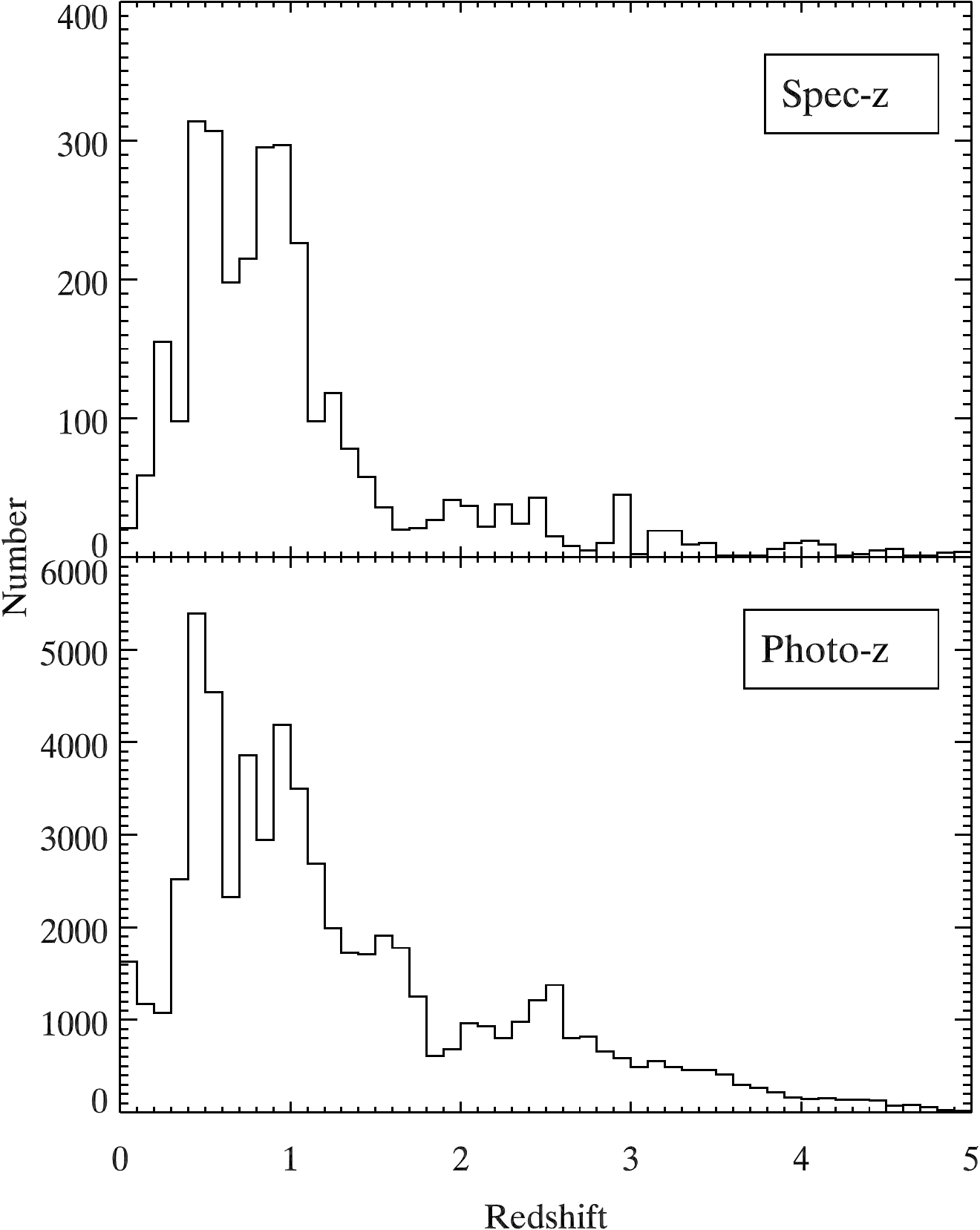}    
   \caption{Spec-$z$ and {\phz} distributions in this work.  
   \label{Fig:phz_spz_hist} }
\end{figure}

Sources with {\spz} (denoted by \zs) information were
used to quantify the {\phz} (denoted by \zp)
performance. Three parameters quantify the {\phz} quality:
$\sigma_{\mathrm{NMAD}}$, $\eta$, and $b_z$. We measured the
normalized median absolute deviation $\sigma_{\mathrm{NMAD}} \equiv
1.48\times \mathrm{median} (\frac { |\Delta z |}{ 1+\zs}$), where
$\Delta z\equiv(\zp-\zs)$. Outliers were not removed before computing
$\sigma_{\mathrm{NMAD}}$. We defined outliers by $\frac{ | \Delta
z|}{1+\zs} > 0.15$, and $\eta$ indicates the fraction of
outliers. $b_z$ is defined by $\frac { \Delta z}{1+\zs}$
which indicates the difference between {\phz} and {\spz}, and
$\overline{b_z}$ is the mean value of $b_z $ after excluding the
outliers. Figure~\ref{Fig:phz_spz_hist} shows the {\phz} and {\spz}
distributions. The majority of sources with spectra have $\zs<1.5$.

\begin{table*}
\begin{center}
\caption{Photo-$z$ quality for non-X-ray and X-ray sources.\label{Table:phz_gal_x}}
\begin{tabular}{ccccc@{\hspace{3em}}cccc}
\tableline
\tableline
  \multicolumn{1}{c}{}&\multicolumn{4}{c}{Non-X-ray sources}&\multicolumn{4}{c}{X-ray sources}\\
  & $N$ &$\overline{b_z}$ & $\sigma_{\mathrm{NMAD}}$ & $\eta(\%)$& $N$ &$\overline{b_z}$& $\sigma_{\mathrm{NMAD}}$& $\eta(\%)$ \\
\tableline
    Total          &2861   &$-0.013$   &0.036    &7.34     &352
    &$-0.003$   &0.040   &10.51 \\ 
\tableline
   $R < 23$    &\0947     &$-0.011$   &0.025    &1.58     &229
   &\00.001   &0.036    &6.99 \\ 
   $R > 23$    &1914   &$-0.014$   &0.044   &10.19   &123
   &$-0.012$   &0.054   &17.07 \\ 
\tableline
  $z < 1.0$     &1893   &$-0.013$   &0.034    &4.07    &224    &\00.001
  &0.035    &5.36 \\ 
  $z > 1.0$     &\0967     &$-0.012 $  &0.042   &13.75    &128
  &$-0.011$   &0.054   &19.53 \\ 
\tableline
\end{tabular}
\end{center}
\end{table*}

\subsection{Non-X-ray sources}

\subsubsection{Galaxy templates}
For the non-X-ray-detected sources, we applied the same templates as
used by \citet{ilbert09} in the Cosmic Evolution Survey (COSMOS)
field. The templates are well-verified SED templates for galaxies
and have been used in many works \citep[e.g.,
][]{salvato09,ilbert2013,hsu14,laigle16}. The library contains 31
templates including 19 galaxies from \citet{polletta07} and 12 young,
star-forming galaxies from \citet{BC03} models.
\textsc{LePhare} adds emission lines to
the galaxy templates during the fitting process (see Section~3.2 of
\citealt{ilbert09}). These lines ([\ion{O}{2}], [\ion{O}{3}],
H$\beta$, H$\alpha$, and Ly$\alpha$) were estimated from fixed
ratios to the UV luminosity as defined by \citet{kennicutt98}. This
helps characterize star-forming galaxies that have
strong emission lines in their SEDs. 

The extinction laws adopted in the fitting were from \citet{prevot84}
and \citet{calzetti00} plus two additional models with modifications
of the 2170\AA\ bump based on the \citeauthor{calzetti00} extinction laws. We allowed
the parameter $E(B-V)$ to range from 0.00 to 0.50 in steps of
0.05~mag. As described by \citet{ilbert09}, we set a prior
of absolute magnitude range to be $-24<M_{B}<-8$ (typical for normal
galaxies) during the fitting procedure to reduce the degeneracy.

Systematic differences can occur in each band between
our photometry and predicted photometry based on the
best-fit template selected during the {\phz} computation. Using the
code \textsc{LePhare}, we derived  zero-point offsets with the
training sample (randomly selected $\sim$800 non-X-ray sources) to
correct the systematic differences for all the sources. The same
corrections were applied to X-ray objects.

\subsubsection{{Photo-$z$} results} \label{sec:gal_phz_result} 
\begin{figure} \centering
  \includegraphics[width=0.45\textwidth]{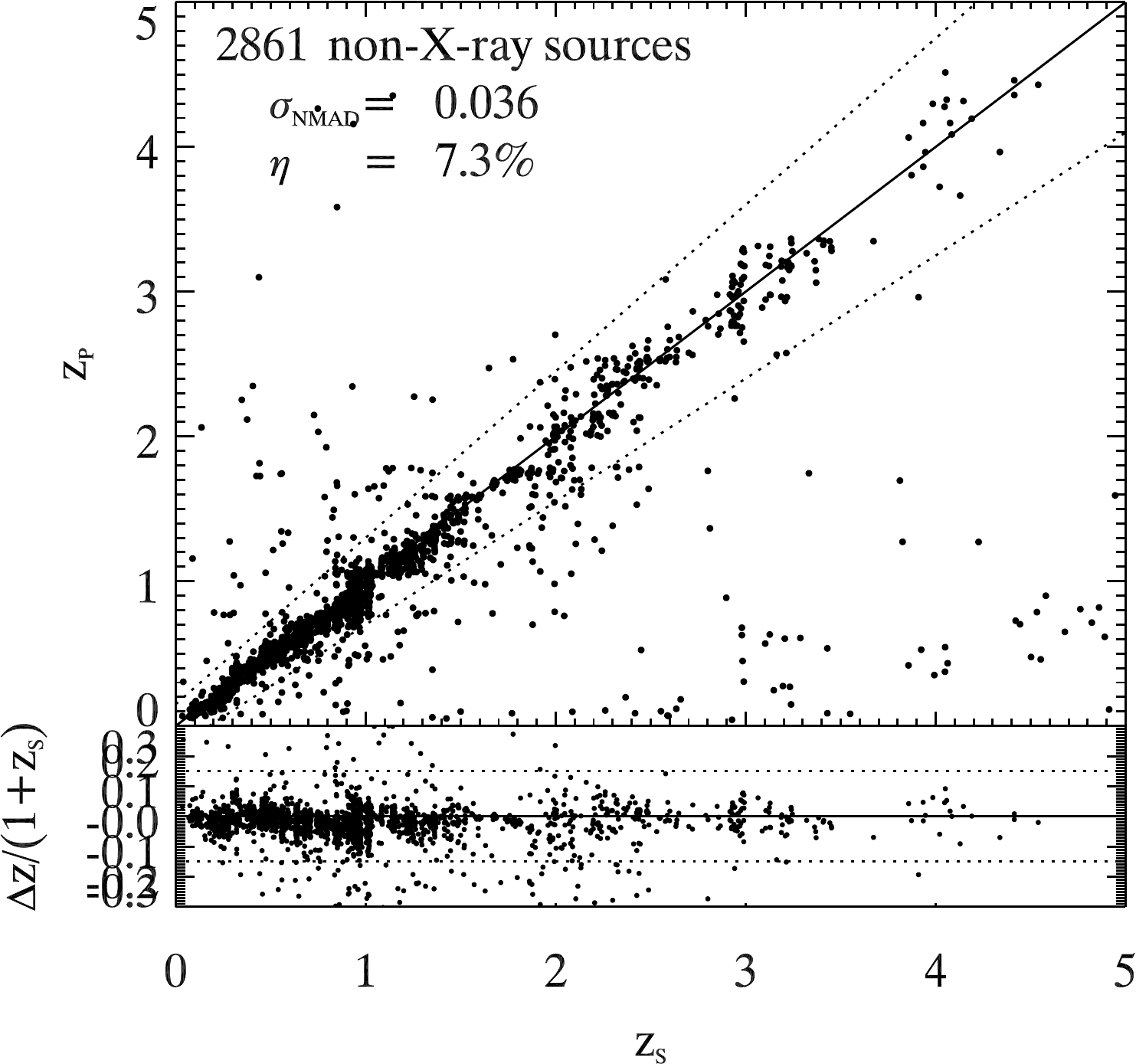}
   \caption{Upper panel: {\spz} vs. {\phz} for non-X-ray sources
(black dots) in this work. The solid line indicates $\zp=
\zs$; the two dotted lines embrace the $\zp=
\zs\pm 0.15(1+\zs)$. Lower panel shows $\Delta z\equiv(\zp-\zs)$. \\
   \label{Fig:spz_vs_phz_gal} }
\end{figure}

The
{\phz} result using 12 bands (i.e., $U$, $B$, $V$, $R$, $I$, $z$,
$y$, $J$, $H$, $K_{s}$, 3.6~and 4.5~\micron) is shown
in Figure~\ref{Fig:spz_vs_phz_gal}. We achieved an overall accuracy
$\sigma_{\mathrm{NMAD}}=0.036$ with an outlier fraction $\eta=7.3\%$
from the comparison between {\phz} and {\spz}. For the bright
($R<23$) sources, we obtained $\sigma_{\mathrm{NMAD}}= 0.025$ and
$\eta=1.58\%$. For low-redshift ($z < 1.0$) sources, we
obtained $\sigma_{\mathrm{NMAD}}=0.034$ and $\eta=4.0\%$
(Table~\ref{Table:phz_gal_x}).

\begin{figure} \centering
  \includegraphics[width=0.23\textwidth]{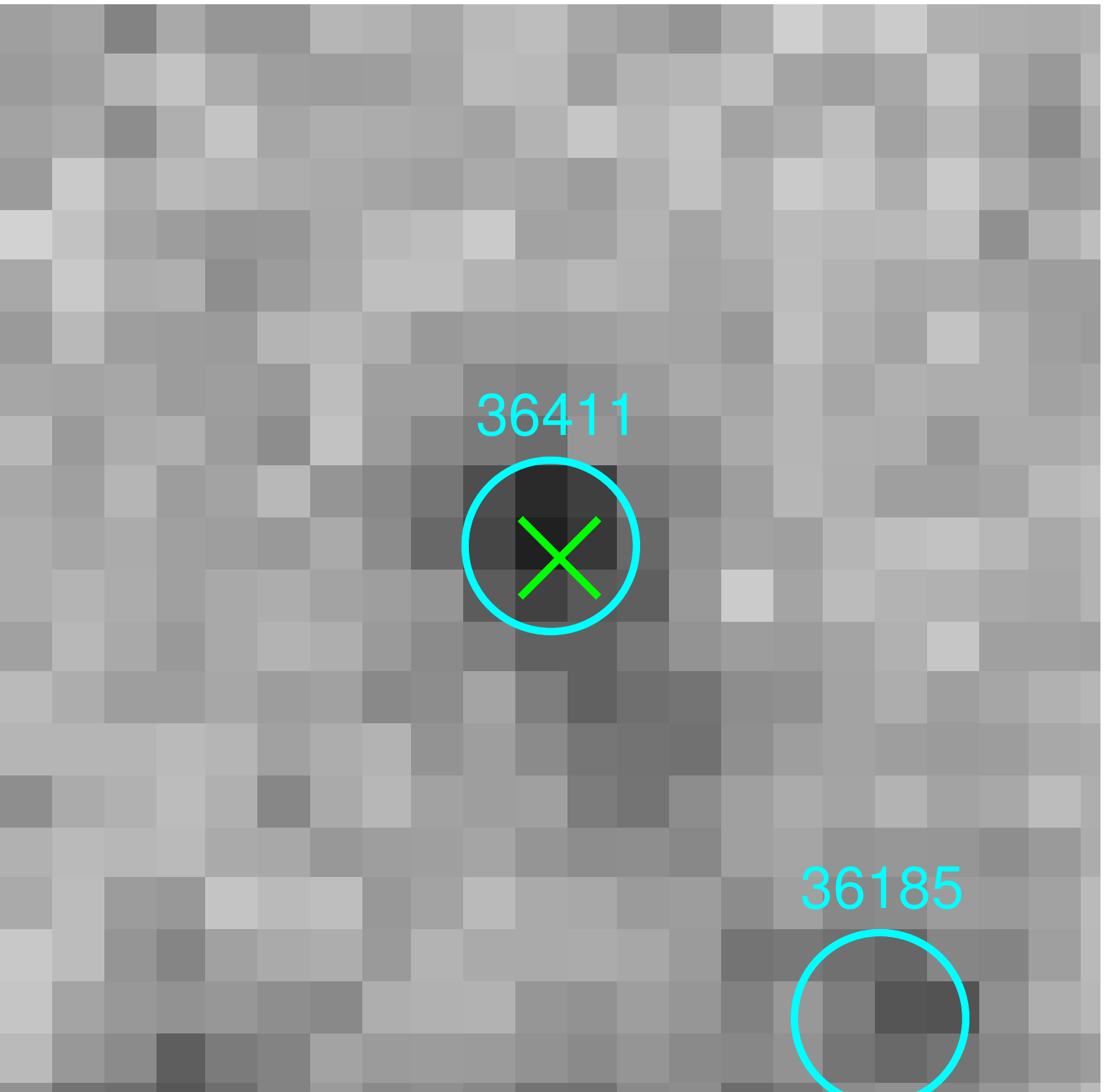}\includegraphics[width=0.23\textwidth]{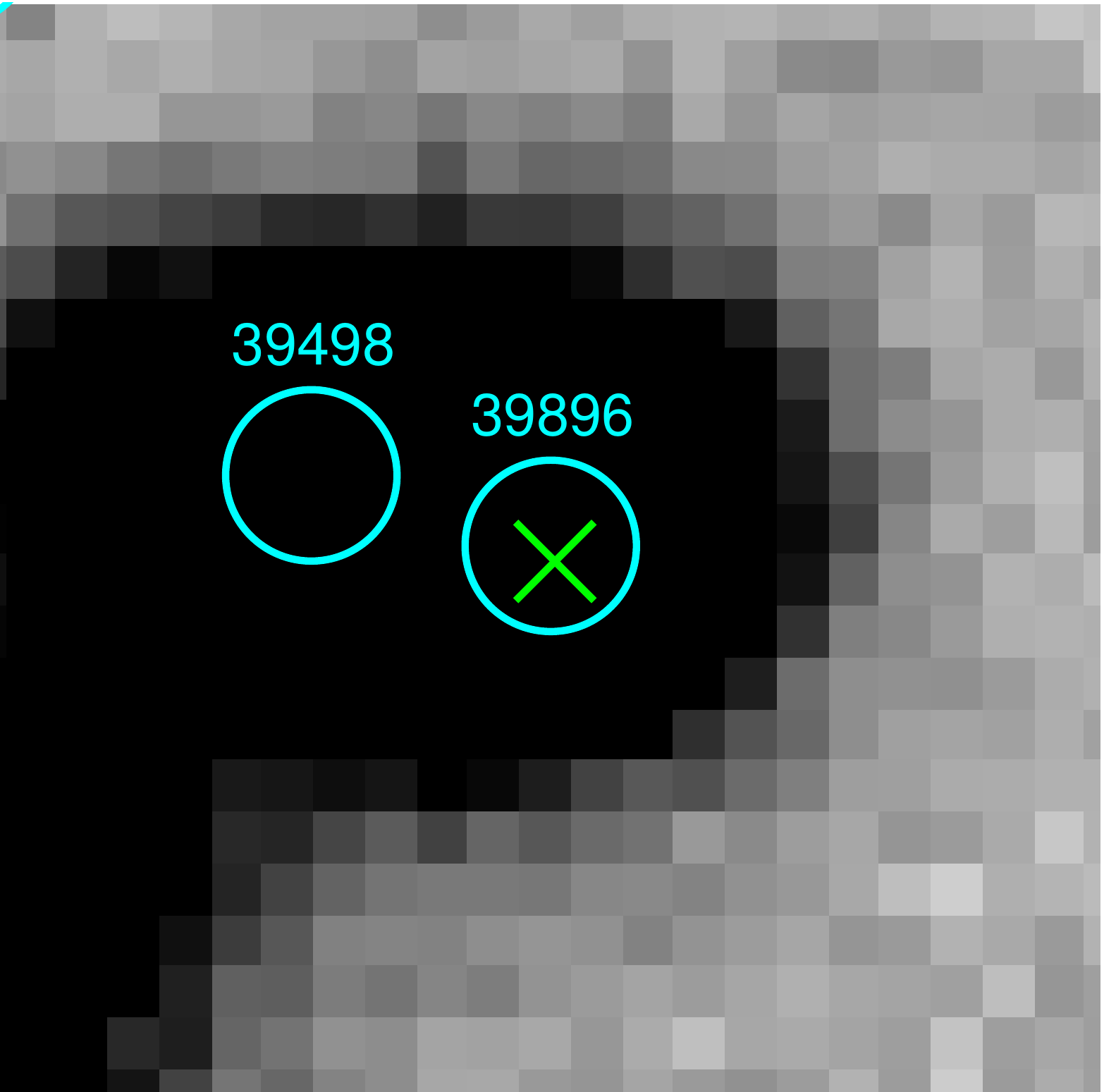}
   \caption{Two examples of {\phz} outliers in $R$-band negative images
6\arcsec$\times$6\arcsec\ in size. Green
crosses mark the {\spz} positions, and cyan open circles 1\arcsec\ in
diameter indicate the
$zJHK_{{s}}$ detections. The left source is faint 
with most of of its visible magnitude $>$25. The right source
is blended, and its photometry is contaminated. 
    \label{Fig:outliers} }
\end{figure}

There are several possible reasons for {\phz} outliers. 
The first is uncertain photometry,
including faint or blended objects. (See two examples in
Figure~\ref{Fig:outliers}.) Second, outliers could be due to the absence of
representative templates in the library. For instance,
\citet{onodera2012} found that {\phz}s in the \citet{ilbert09} sample
are often underestimated for 
quiescent galaxies at $1.5<z<2$. Third, 
misinterpreting a spectral break as Lyman break or
Balmer break could lead to a catastrophic failure of {\phz}
\citep{dahlen2010}.
Fourth, the broad-band data points may miss or mis-fit 
emission lines in the SEDs. In this case, as demonstrated by
\citet{ilbert09} and \citet{salvato09}, medium- or narrow-band data
can help to pinpoint the emission line features and significantly
improve the {\phz} quality. Due to the lack of medium- or
narrow-band data in this work, it is not surprising that we are not
able to achieve the same {\phz} qualities as presented in other deep
fields such as COSMOS and CDFS. The upcoming Survey for High-$z$
Absorption Red and Dead Sources \citep[SHARDS;] []{shards2013} will
help to improve the {\phz} quality in GOODS-N region.

\subsection{X-ray sources} \label{X_analysis} Because the majority of
X-ray detections are associated with AGNs, we cannot use pure galaxy
templates to fit their SEDs. A separate set of AGN-galaxy hybrid
templates is required.

\subsubsection{Cross-matching}\label{Cross-matching} 
The first step in computing {\phz}s for X-ray sources was to cross-match
our $zJHK_{{s}}$ detections to the 2Ms-CDFN X-ray catalog 
\citep{xue16}. The X-ray source positions used $K_{s}$-band
images from \citet{wang10a} as the reference frame, which is the same
as ours. The median offset between the X-ray positions and our
$zJHK_{{s}}$ positions is
0\farcs2. This value is smaller than
the median value of X-ray positional uncertainty
(0\farcs6). Therefore, we adopted a
simple search radius  of 1\arcsec\ to match our $zJHK_{{s}}$
detections with X-ray sources. 602 X-ray sources have $zJHK_{{s}}$
detections within this search radius.
\citet{xue16} used likelihood-ratio matching to identify the
visible/near-infrared/mid-infrared/radio (ONIR) counterparts.
We compared their positions to those of our $zJHK_{{s}}$
counterparts. When more than one $zJHK_{{s}}$ detection was found 
within the search radius, we chose the one closest to the
ONIR counterpart. $zJHK_{{s}}$ counterparts within
1\arcsec\ radius are labeled ``xflag=1'' in the
catalog.

\begin{figure} \centering
  \includegraphics[width=0.5\textwidth]{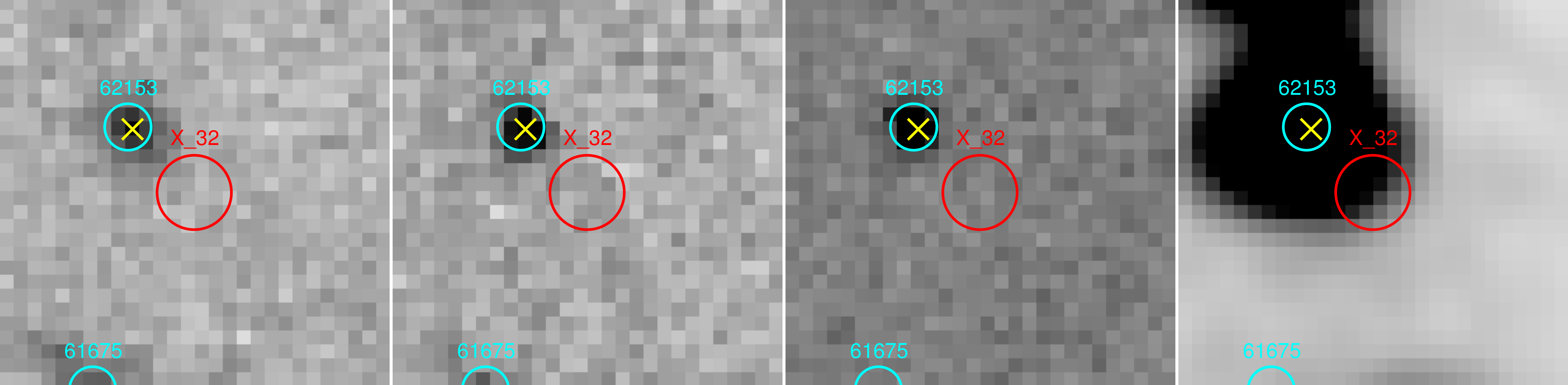}
   \caption{Example of multi-band cutouts for one X-ray source with
$zJHK_{{s}}$ counterpart separated by more than
1\arcsec\ from the \citet{xue16} ONIR counterpart position. From left
to right, negative images show $R$, $z$, $H$, and IRAC
3.6~$\mu$m. Images are 8\arcsec$\times$8\arcsec\ on a side. Yellow
crosses mark the ONIR counterparts 
found by \citet{xue16}, cyan open circles mark the corresponding
$zJHK_{{s}}$ detection from this work, and red open circles mark
the X-ray positions with radius being the positional uncertainty. In
this case, the separation between the X-ray source ([XLB2016 CDFN]
32) and the $zJHK_{{s}}$ detection is about 2\arcsec.
   \label{XID_32} }
\end{figure}

For the remaining 81 X-ray sources lacking counterparts within the
1\arcsec\ search radius, we checked their multi-band cutouts
visually. 56 of them have $zJHK_{{s}}$ detections nearby that are
consistent with the ONIR counterparts from \citet{xue16}, as shown in
Figure~\ref{XID_32}. In our released catalog, we treated these 56
sources as the $zJHK_{{s}}$ counterparts of the X-ray sources and
provided {\phz} measurement of them. The {\phz}s of these 56
sources should be used with caution because of the large separation
as shown in Figure~\ref{XID_32}. These sources are labeled
``xflag=2'' in the catalog.

\begin{figure} \centering
          \includegraphics[width=0.5\textwidth]{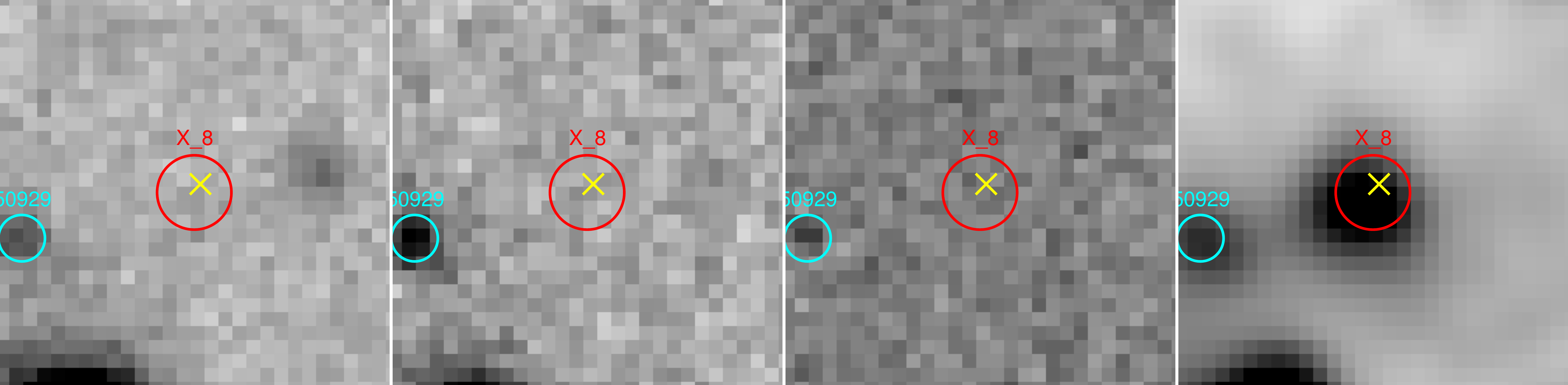}
\\ 
          \includegraphics[width=0.5\textwidth]{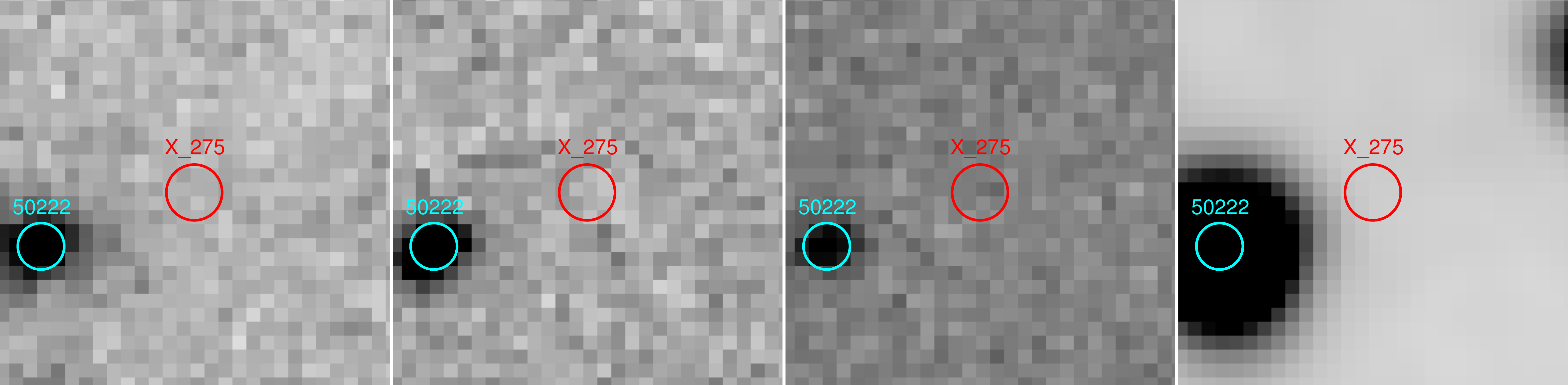}
\\ 
         \includegraphics[width=0.5\textwidth]{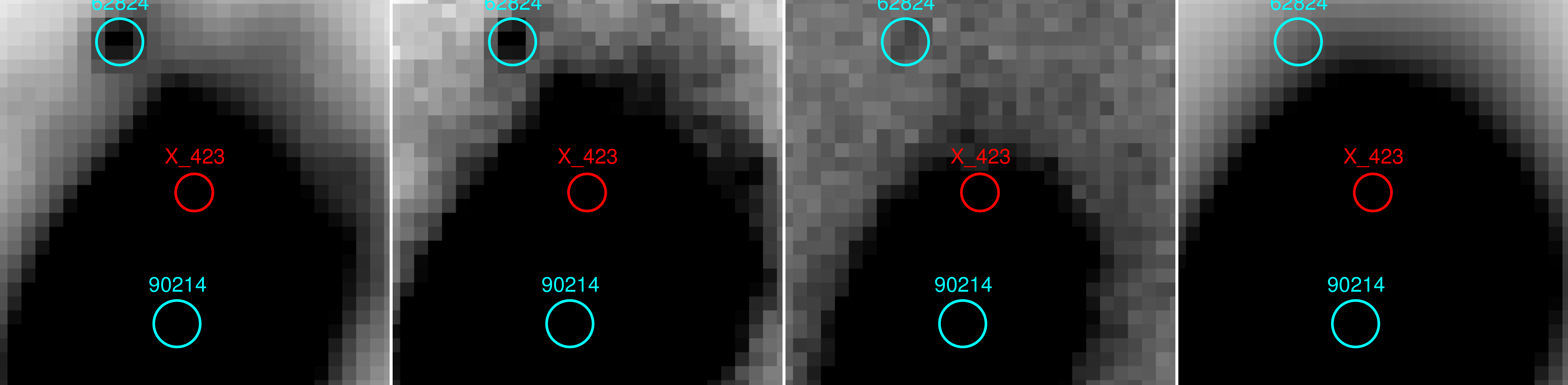}
\caption{Three examples of multi-band cutouts for X-ray sources
without $zJHK_{{s}}$ counterparts. From left to right are $R$,
$z$, $H$, and IRAC 3.6~$\mu$m negative images 8\arcsec$\times$8\arcsec\ on a
side. Yellow crosses mark the ONIR counterparts 
\citep{xue16}, cyan open circles mark the corresponding
$zJHK_{{s}}$ detections from this work, and red open circles mark
the X-ray positions with radius being the positional
uncertainty. From top to bottom are [XLB2016 CDFN] 8, 275, and 423
respectively.\label{Xctp}}
 \end{figure}

In total, 25 out of 683 X-ray sources have no $zJHK_{{s}}$
counterparts. With visual inspection, we group these 25 sources into
three cases:
\begin{itemize}
   \item Case~1: 8 sources that are only detected by IRAC
(Figure~\ref{Xctp}a);
   \item Case~2: 13 sources that have no $zJHK_{{s}}$ or IRAC
detections around the X-ray detection (Figure~\ref{Xctp}b);
   \item Case~3: 4 sources that have highly extended $zJHK_{{s}}$
sources nearby. These sources have a large offset to the X-ray source
position and therefore are not likely counterparts
(Figure~\ref{Xctp}c).
\end{itemize}

In Case~1 above, the IRAC sources can be
considered the correct identifications for the X-ray
sources. However, because Case~1 has only IRAC photometry, we cannot
compute {\phz}s for these eight sources. In Case~2, 11 of them have
either no $zJHK_{{s}}$ or ONIR counterparts around the X-ray
detection, whereas the remaining two X-ray sources have ONIR
counterparts in bluer bands but no $zJHK_{{s}}$ detections.
For Case~3, the extended $zJHK_{{s}}$ sources are likely
foreground galaxies that hide the true counterpart. 
We have excluded objects in all three cases above from the subsequent
analysis for the X-ray sources.

\subsubsection{AGN-galaxy hybrid training} After cross-matching, we
computed {\phz}s for the X-ray sources in the CDFN. We applied the
AGN-galaxy hybrid templates built by \citet{hsu14}, which were
trained on the X-ray sources detected in the 4Ms-CDFS survey. Each
AGN-galaxy hybrid is composed of a galaxy template and an AGN
template with ratios varying from 1:9 to 9:1. The galaxy templates
are from \citet{bender01}, and the AGN templates are from
\citet{polletta07}. The best set of hybrid templates were tuned using
randomly-selected $25\%$ of the X-ray sources from the 4Ms-CDFS
catalog. The AGN-galaxy hybrid libraries used here contain 48 and
30 best-fit templates for optically extended sources and point-like
sources, respectively. Details were given by \citet{hsu14}.

During the SED fitting process for the X-ray sources, we adopted the
same systematic offsets and extinction laws as used for the non-X-ray
sources. For an absolute magnitude prior, we used $-24<M_{B}<-8$ for
extended X-ray counterparts and $-30<M_{B}<-20$ for point-like X-ray
counterparts \citep{salvato09}. 

\subsubsection{Photo-$z$ results}

\begin{figure} \centering
  \includegraphics[width=0.45\textwidth]{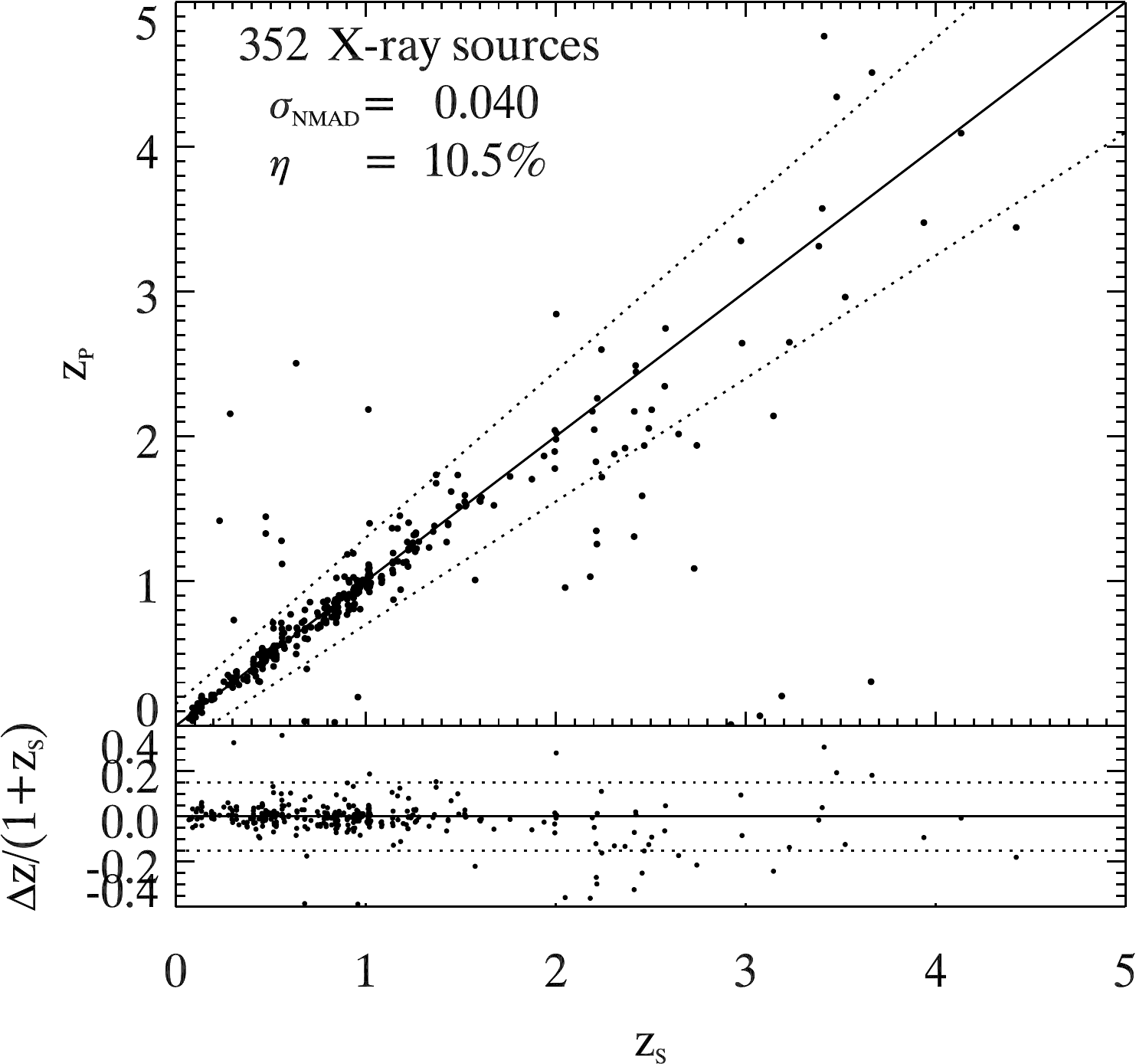}
   \caption{Spec-$z$ vs. {\phz} for all the X-ray sources (black dots)
with {\spz} in this work. The solid line indicates $\zp=
\zs$; the two dotted lines embrace the $\zp=
\zs\pm 0.15(1+\zs)$. Lower panel shows $\Delta z\equiv(\zp
-\zs)$.  \label{Fig:spz_vs_phz_x} }
\end{figure}

\begin{figure} \centering
  \includegraphics[width=0.45\textwidth]{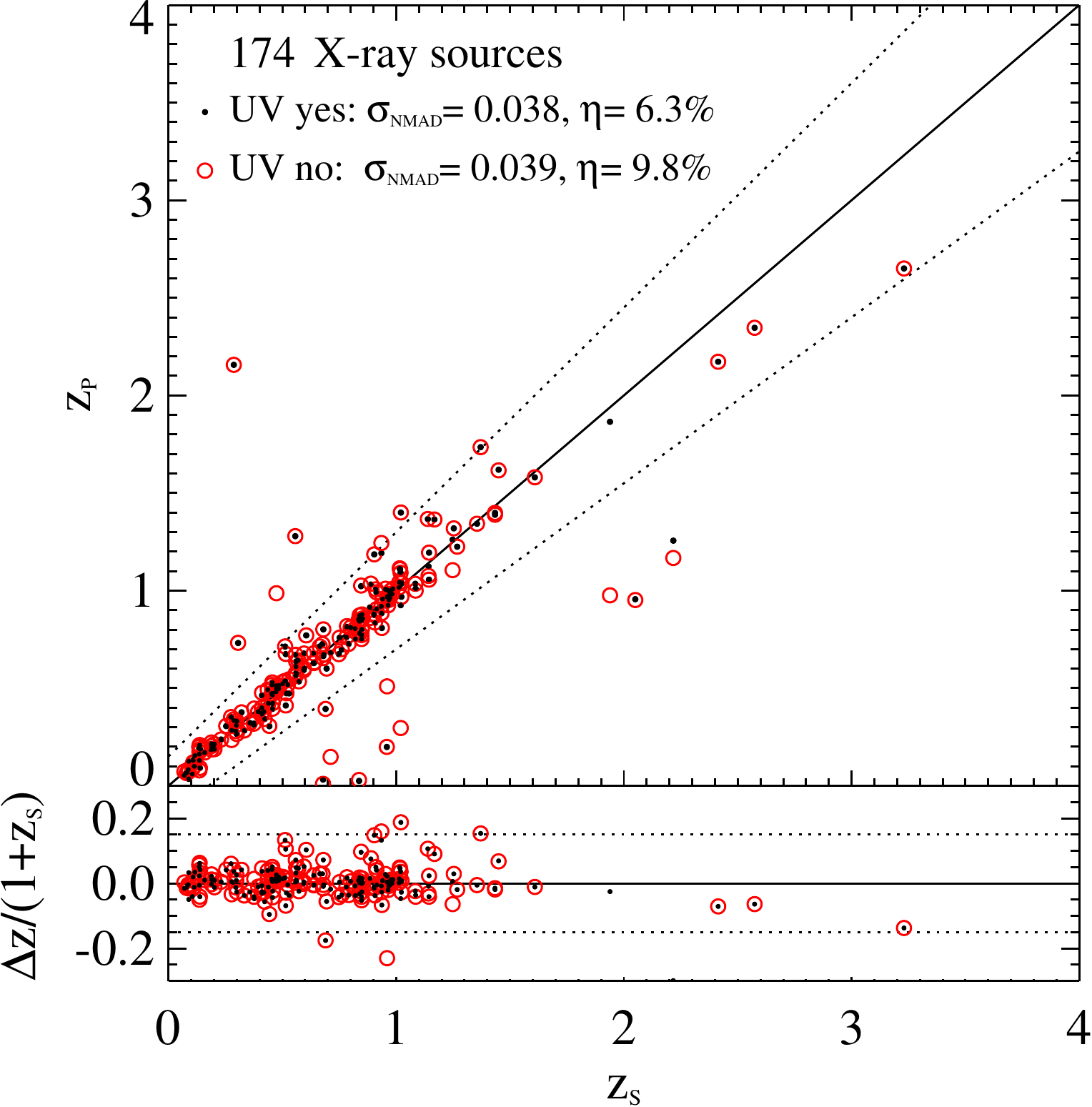}
   \caption{Spec-$z$ vs. {\phz} for 174 X-ray sources that have either
FUV or NUV data. The black dots indicate the {\phz} computed with UV
data, and the red open circles indicate the {\phz} computed without
UV data. \label{Fig:ECDFN_xray_UV} }
\end{figure}

The UV bump from an AGN accretion disk is a significant spectral
feature for distinguishing an AGN from a non-active
galaxy. Therefore, UV data are particularly important for obtaining
accurate {\phz}s for AGNs, as demonstrated by \citet{hsu14}. 222 out
of 683 X-ray sources are detected in either FUV or NUV. For these
sources, we included UV data in addition to the same 12 bands used
for the non-X-ray objects when computing {\phz}s.
As shown in Table~\ref{Table:phz_gal_x} and Figure~\ref{Fig:spz_vs_phz_x}, we achieved an overall
$\sigma_{\mathrm{NMAD}}$  for X-ray counterparts almost as good as
that for non-X-ray sources. The outlier fraction was higher, especially
for fainter sources and those at $z>1$. To demonstrate the benefit of UV
data, we compared {\phz} results with and without
UV data for 174 X-ray sources that have {\spz}.
Figure~\ref{Fig:ECDFN_xray_UV} shows that
$\sigma_{\mathrm{NMAD}}$ is almost unchanged, but 
the outlier fraction $\eta$ decreases from $9.8\%$ without UV data to
$6.3\%$ when UV data are added.

The causes of the catastrophic failures of {\phz}s for the X-ray
sources are more complicated than for the non-X-ray sources. In addition
to the reasons discussed in Section~\ref{sec:gal_phz_result}, the
outliers could be also due to variability \citep{salvato09} and to the uncertain
contributions from AGNs. 
  
\begin{figure*} \centering
   
          \includegraphics[width=\textwidth]{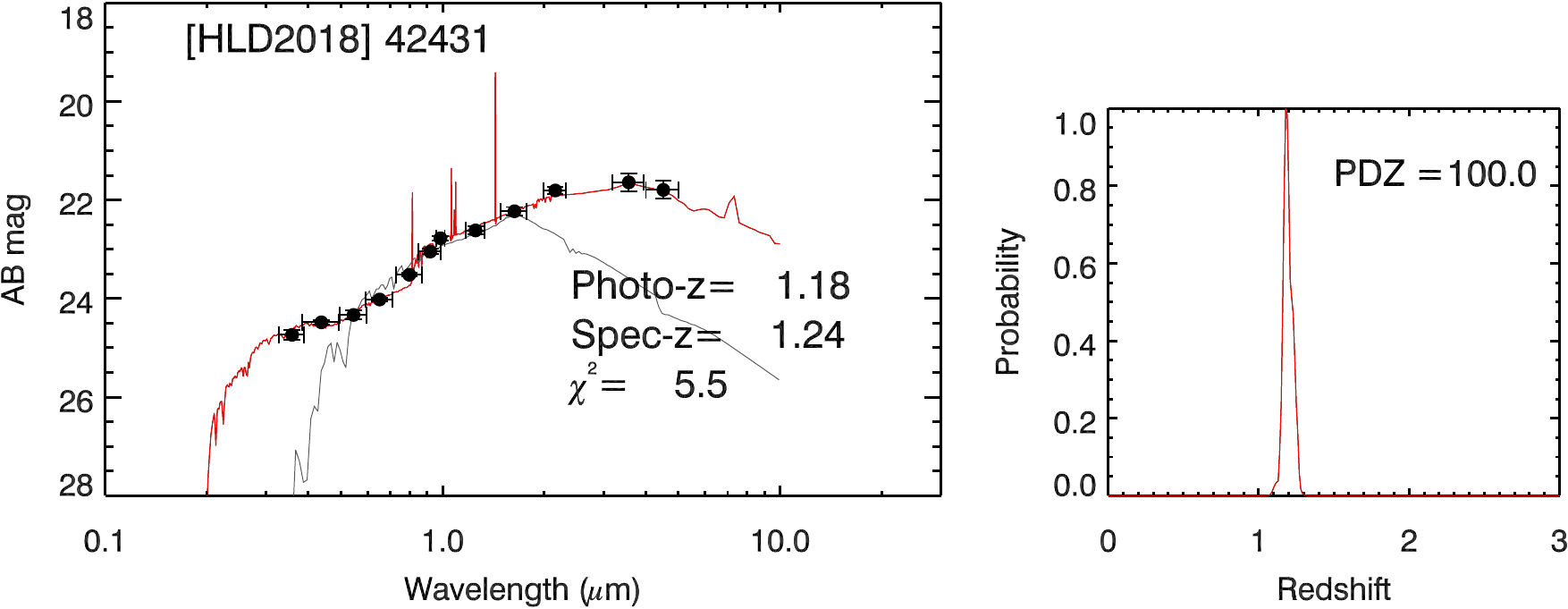}
  \label{42431}
 
          \includegraphics[width=\textwidth]{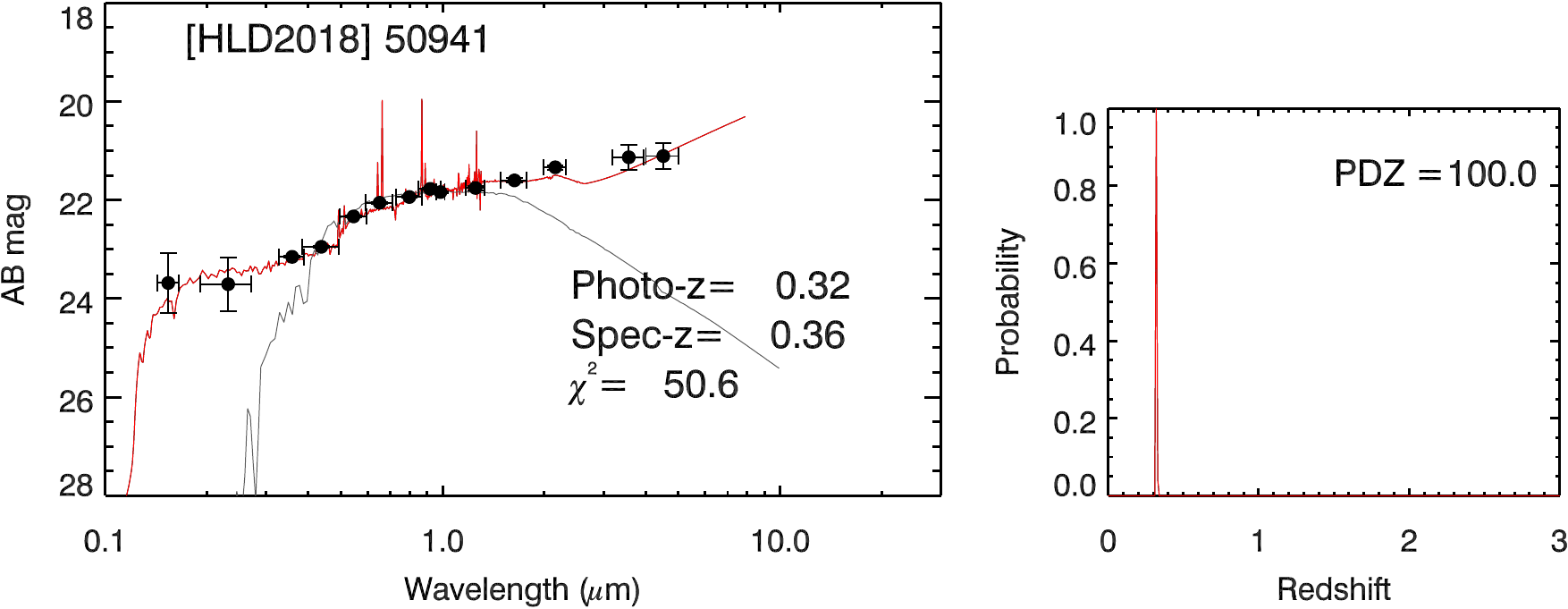}
  \label{50941}
 
          \includegraphics[width=\textwidth]{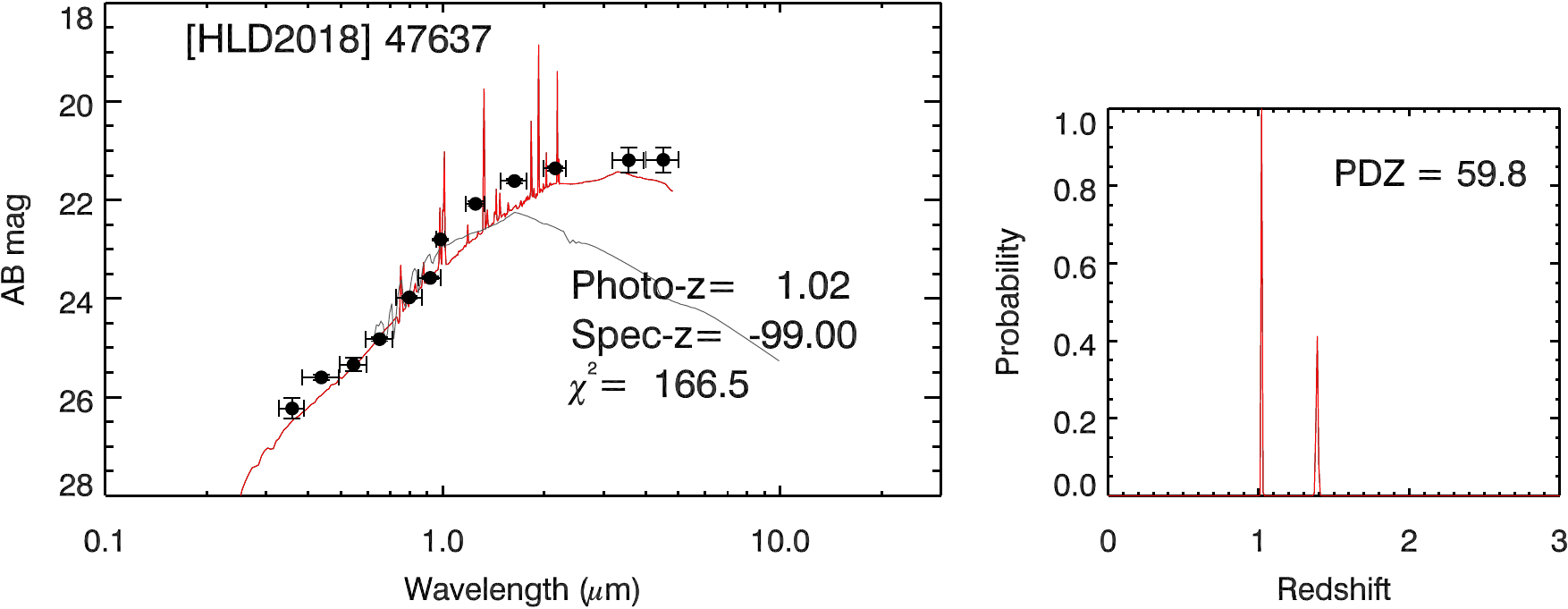}
  \label{47637}
 
\caption{Examples of SED fitting. Left panels show the observed data
points (black filled circles) with the best-fit templates (red solid
line) and stellar templates (grey solid line). Right panels
show the corresponding redshift probability distribution function
$p(z)$. The sources from top to bottom are [HLD2018] 42431, 50941,
and 47637, respectively. The first one is a non-X-ray source fitted
with a galaxy template. The two below are X-ray sources fitted with
AGN--galaxy hybrid templates. The PDZ is the normalized integration of $p(z)$ between
$\zp\pm 0.1(1 + \zp)$. Higher values of PDZ
indicates higher probability of the given
{\zp}.  \label{SEDfit} }
\end{figure*}

In our released catalog, for both X-ray and non-X-ray sources,
{\zp} is defined by the peak value of the redshift
probability distribution function $p(z)$. Figure~\ref{SEDfit} shows
three examples of our  results. In some cases (e.g., the
object [HLD2018]=47637 in Figure~\ref{SEDfit}), there
are two or more peaks in the probability distribution function.
In such situations, we defined {\phz} as the highest peak of
$p(z)$. 

\subsection{Comparison with previous work}

\begin{figure} \centering
  \includegraphics[width=0.45\textwidth]{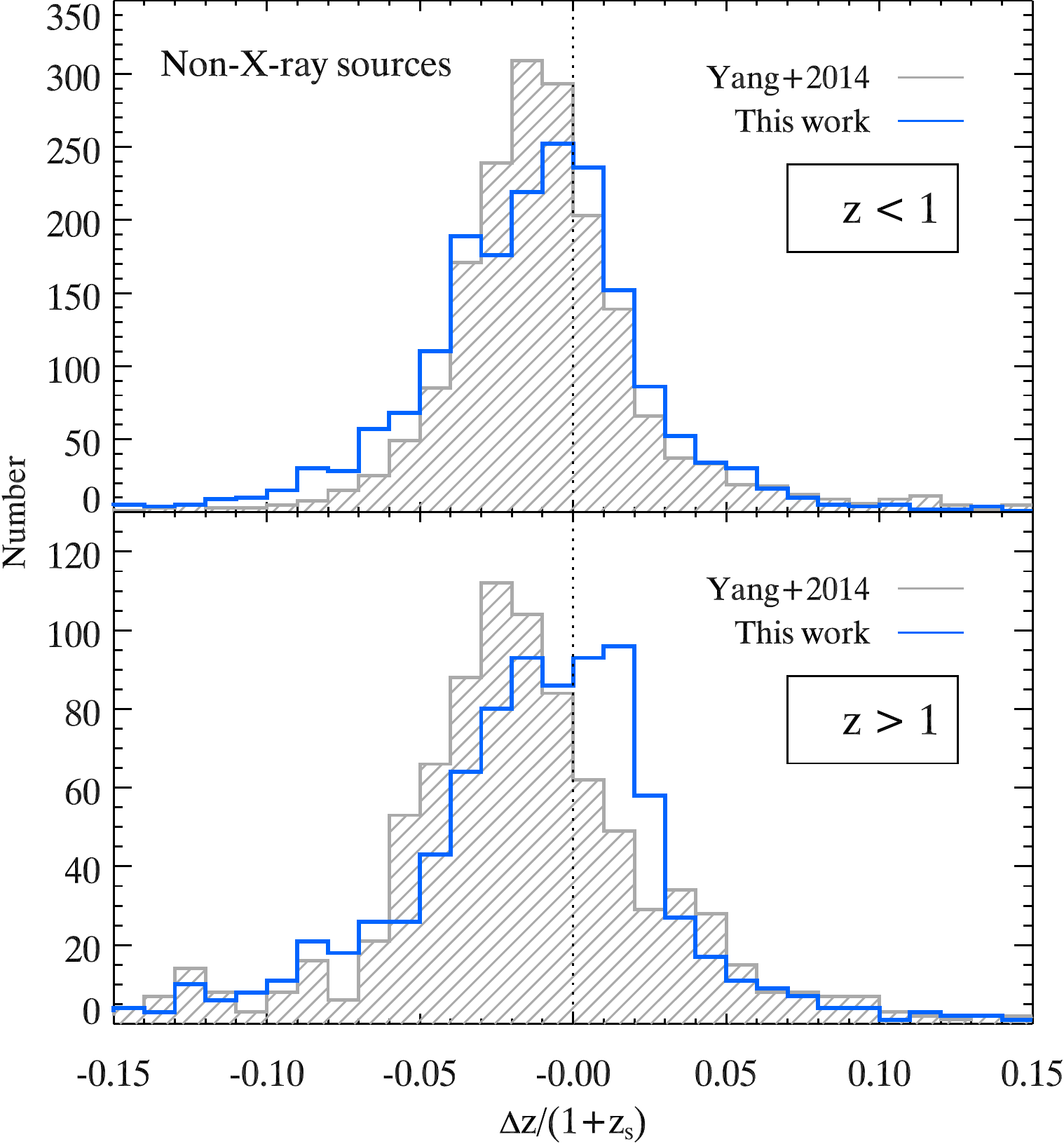}
  \caption{Comparison of the $\frac { \Delta z}{1+\zs}$
distribution for the non-X-ray sources with the previous work of
\citealt{yang14} at $z<1$ (upper panel) and $z>1$ (lower
panel). \label{Fig:LTvsYang_hist_zbin} }
\end{figure}

\citet{yang14} used the code \textsc{EAzY}
\citep{brammer2008} to estimate {\phz}s for sources in the
field. The main differences between our work and theirs are:
(1) the templates used for SED fitting. \citet{yang14} used
only 8 galaxy templates for non-X-ray sources and added 3 QSO
templates for X-ray sources. We used 31 galaxy templates for
non-X-ray sources and 78 galaxy or galaxy--AGN hybrid templates
for X-ray sources. (2) the integration time of the NIR images. The
NIR data we used are $0.5-1.0$~mag deeper. (3) the
X-ray catalog. \citet{yang14} used the
\citet{alexander03} X-ray catalog, whereas we used the one
from \citet{xue16}. (4) UV data. We added UV data to compute
{\phz}s for the X-ray sources.
For 2842 non-X-ray sources with {\spz}s
in common between our work and
\citet{yang14}, the two studies achieved  comparable quality
($\sigma_{\mathrm{NMAD}} = 0.036$, $\eta = 7.1\%$).
At $z<1$, both works have
similar error distributions as shown in Figure~\ref{Fig:LTvsYang_hist_zbin}. At $z>1$, our work is more peaked near
$\frac { \Delta z}{1+\zs}=0$ and has smaller 
scatter ($\sigma_{\mathrm{NMAD}}=0.042$), whereas \citet{yang14}
tended to underestimate {\phz}s and had slightly larger scatter
($\sigma_{\mathrm{NMAD}}=0.048$).

\begin{figure} \centering
  \includegraphics[width=0.45\textwidth]{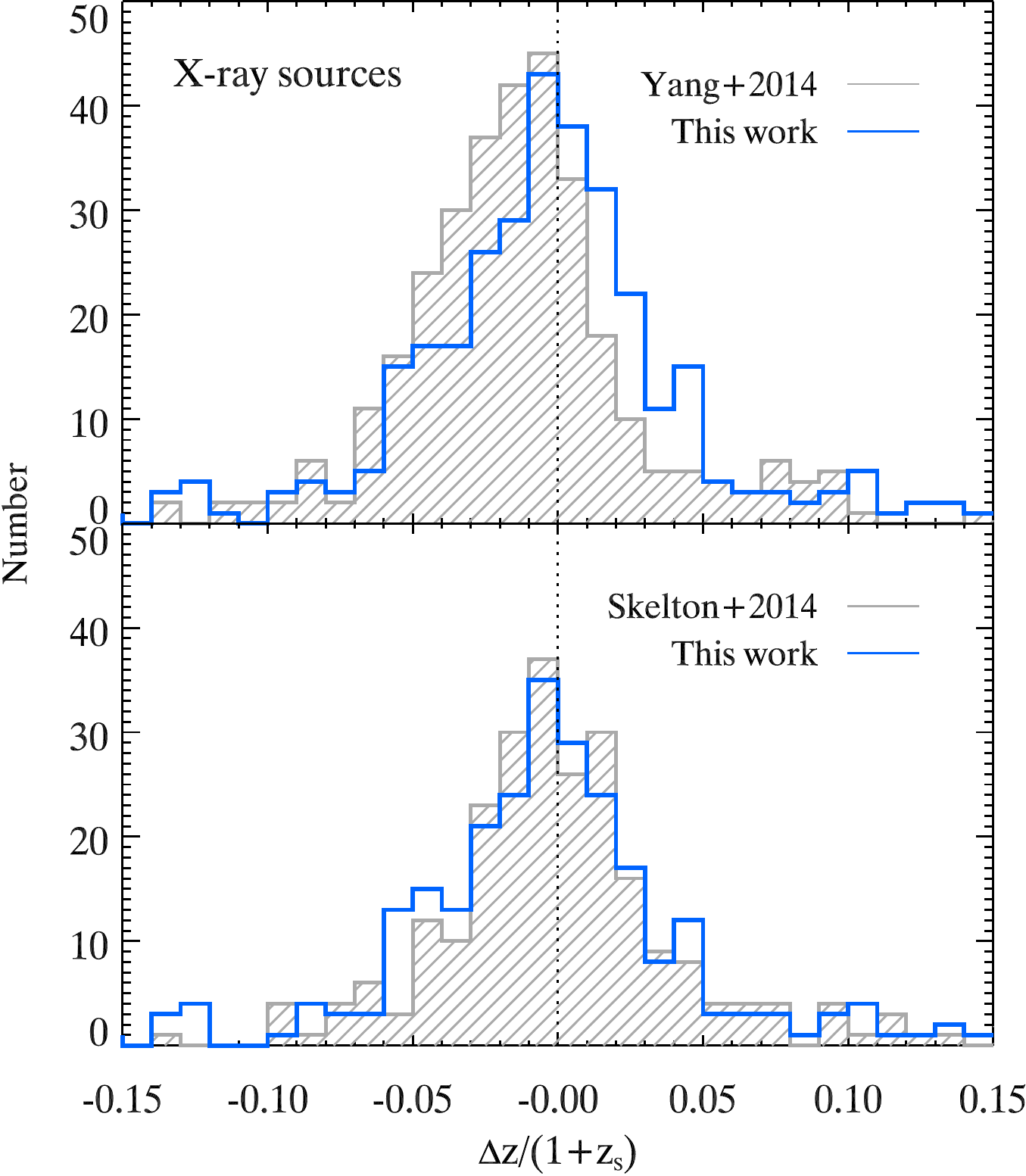}
  \caption{Comparison of the $\frac { \Delta z}{1+\zs}$
distribution for the X-ray sources with the previous
studies of \citealt{yang14} (upper panel) and \citealt{skelton14}
(lower panel). \label{Fig:LTvsYang_hist_x} }
\end{figure}

For the X-ray sources, \citet{xue16} adopted two main \phz\
catalogs \citep{yang14,skelton14}.
Table~\ref{Table:phz_x_HvsY} lists the comparisons for
the 269 sources in common among all three works. Our {\phz}
accuracy is similar to that of
\citet{yang14}. The accuracy of the 3D-HST catalog
\citep{skelton14} is better than the other two studies for both X-ray
and non-X-ray sources because
\citet{skelton14} used the extremely deep visible and NIR data from
{\it HST}. However, \citet{skelton14} fitted  the  X-ray
sources with normal galaxy templates, and that gave  a larger
outlier fraction ($11.9\%$) compared to ours ($8.6\%$).
Figure~\ref{Fig:LTvsYang_hist_x} compares our $\frac { \Delta z}{1+\zs}$ distribution
with the two other studies. Although \citeauthor{yang14} achieved
slightly lower outlier
fraction than we did, their $\frac { \Delta
z}{1+\zs}$ distribution for X-ray sources has a negative
offset (see the upper panel in Figure~\ref{Fig:LTvsYang_hist_x}),
which was also mentioned by \citet{xue16}. This is probably caused by
systematic errors. The offsets of $\frac{\Delta z}{1+\zs
}$ are quantified by $\overline{b_z}$ in
Table~\ref{Table:phz_x_HvsY}.

\section{Column description for the released
catalogs} \label{column_description}

We release photometry and {\phz} catalog as shown in
Table~\ref{cat:photometry}. We also release the NIR images online.
All data are available through the portal: {\url
http://idv.sinica.edu.tw/lthsu}. Below are the descriptions for the
table columns.
\begin{itemize}
\item Column~1 ([HLD2018]) gives the sequence number of
$zJHK_{{s}}$ detections.
\item Column~2 (R.A.) gives the J2000 right ascension of
$zJHK_{{s}}$ detections.
\item Column~3 (Decl.) gives the J2000 declination of
$zJHK_{{s}}$ detections.
\item Column~4 (\zs) gives the spectroscopic
redshift. The value $-99$ indicates that no {\zs} is  available.
\item Column~5 ($Q_{{s}}$) gives the spectroscopic redshift
quality. The value 1 indicates that the {\zs} is good, and the value 2
indicates that the {\zs} is uncertain. The value $-99$ indicates that
no value is  available.
\item Column~6 (\zp) gives the best-fit {\phz}, which is
the highest peak in the associated redshift distribution function
$p(z)$. (See the examples in 
Figure~\ref{SEDfit}.) The value $-99$ indicates that no {\zp} is
available.
\item Column~7 ($1\sigma^{\rm{low}}$): lower 1$\sigma$ value of
{\zp}  estimated from the equation $\chi^2(z) =
\chi^2_{\rm{min}} +1$, where $ \chi^2_{\rm{min}}$ is the minimum of
$\chi^2$ \citep{ilbert09}.
\item Column~8 ($1\sigma^{\rm{up}}$): upper 1$\sigma$ value of {\zp}.
\item Column~9 ($\chi^2_{\rm{best}}$): $\chi^2$ for the best
galaxy fit for non-X-ray sources or AGN--galaxy fit for X-ray sources.
\item Column~10 ($\chi^2_{\rm{star}}$): $\chi^2$ for the best stellar
fit.
\item Column~11 (PDZ): a measure of the probability of \zp: the
integral of the normalized probability distribution function
$p(z)$ between $\zp\pm 0.1(1 + \zp)$. A higher
value of PDZ indicates higher probability of the given {\zp}.
\item Column~12 (xflag):  marks whether a
$zJHK_{{s}}$ source is considered as the counterpart of a given
X-ray source. xflag = 1 indicates a $zJHK_{{s}}$ source  separated
by less than 1\arcsec\ from an X-ray source. xflag = 2 indicates a
$zJHK_{{s}}$ separated from an X-ray source ${>}$1\arcsec.
$\mathrm{xflag}=-99$ indicates $zJHK_{{s}}$ source not considered
a counterpart of any X-ray source. Section~\ref{Cross-matching}
describes the counterpart identification process.
\item Column~13 ([XLB2016 CDFN]) is the sequence number for the X-ray
source given from the 2Ms-CDFN X-ray catalog \citep{xue16}. The value
$-99$ indicates that the source has no X-ray counterpart.
\item Column~14 (XRA) is the J2000 right ascension of the X-ray
source in the 2Ms-CDFN X-ray catalog. The value $-99$ indicates
that the source has no X-ray counterpart.
\item Column~15 (XDEC) is the J2000 declination of the X-ray source
in the 2Ms-CDFN X-ray catalog. The value $-99$ indicates that
the source has no X-ray counterpart.
\item Columns~16--19 give the total flux densities and uncertainties for FUV
and NUV queried from {\it GALEX} GR6/7.
\item Columns~20--39 give the 2\arcsec\ diameter aperture flux densities and
associated uncertainties from \textsc{SExtractor} output parameters
\texttt{FLUX\_APER} and
\texttt{FLUXERR\_APER} for $U$, $B$,
$V$, $R$, $I$, $z$, $y$, $J$, $H$, $K_{{s}}$, respectively.
Each flux density is immediately followed by its uncertainty.
Unit of flux density is $\mu$Jy .
\item Columns~40--59 give the flux densities and associated uncertainties
 from \textsc{SExtractor} output parameters
\texttt{FLUX\_AUTO} and
\texttt{FLUXERR\_AUTO} for $U$, $B$, $V$, $R$, $I$, $z$, $y$, $J$, $H$,
$K_{{s}}$, respectively.
\item Columns~60--63 give the total flux densities and uncertainties for IRAC
3.6~and 4.5~$\mu \rm{m}$ based on \texttt{IRACLEAN}.

\end{itemize}

\begin{table*}
\begin{center}
\caption{Comparison of our X-ray-source {\phz} quality with recent work \label{Table:phz_x_HvsY}}
\begin{tabular}{ccccccccccc}
\tableline
\tableline
  \multicolumn{2}{c}{}&\multicolumn{3}{c}{This work}
  &\multicolumn{3}{c}{\citet{yang14}}
  &\multicolumn{3}{c}{\citet{skelton14}}\\ 
  \cmidrule(r){3-5} \cmidrule(r){6-8}  \cmidrule(r){9-11}
  & $N$ &$\overline{b_z}$ & $\sigma_{\mathrm{NMAD}}$ & $\eta(\%)$ &$\overline{b_z}$& $\sigma_{\mathrm{NMAD}}$& $\eta(\%)$ &$\overline{b_z}$& $\sigma_{\mathrm{NMAD}}$& $\eta(\%)$ \\
\tableline
    Total           &269   &-0.003   &0.039   &8.55   &-0.014
    &0.040    &7.43    &0.001   &0.032   &11.90\\ 
\tableline
   $R < 23$      &170    &0.000   &0.036    &4.12   &-0.016   &0.035
   &2.94    &0.002   &0.030    &7.06\\ 
   $R > 23$       &99   &-0.011   &0.055   &16.16   &-0.009   &0.053
   &15.15   &-0.002   &0.042   &20.20 \\ 
\tableline
  $z < 1.0$       &169   &-0.001   &0.037    &3.55   &-0.015   &0.037
  &4.73    &0.005   &0.030    &5.33 \\ 
  $z > 1.0$       &100   &-0.007   &0.053   &17.00   &-0.012   &0.048
  &12.00   &-0.008   &0.043   &23.00 \\ 
\tableline
\end{tabular}
\end{center}
\end{table*}

\begin{table*}\footnotesize
\centering
\caption{Released photometric catalog\label{cat:photometry}}
\begin{tabular}{lllllllllllr} 
\tableline
\tableline
[HLD2018] &R.A. &Decl. &\zs &$Q_{\rm{spec}}$ &\zp&$1\sigma^{\rm{low}}$&$1\sigma^{\rm{up}}$ & $\chi^2_{\rm{best}}$ & $\chi^2_{\rm{star}}$ & PDZ  &xflag \\
(1) &(2) &(3) &(4) &(5) &(6) &(7) &(8) &(9) &(10) &(11) &(12)  \\
\tableline
18637 & 189.264864  & 62.0859387  & $-99$        &$-99$   & 0.9024 &
0.8956 & 0.9104 & 44.6636 & 429.492 & 100    &1 \\ 
18709 & 189.1442586 & 62.0872443 & 0.6346  & 1     & 0.6243 & 0.616 &
0.6328 & 16.4447 & 731.437 & 100  & $-99$  \\ 
18837 & 189.3097146 & 62.0846307 & $-99$        & $-99$  & 0.423 & 0.4187
& 0.4255 & 75.3598 & 3006.53 & 100  &1\\ 
18864 & 189.2370479 & 62.0831875  & $-99$       & $-99$  & 0.8087 & 0.794
& 0.8219 & 91.5799 & 903.539 & 100  &1 \\ 
19099 & 189.1462737 & 62.0888751 & 0.9876   & 1     &0.9252 & 0.9181
& 1.0144 & 36.7712 & 4045.29 & 100  & $-99$ \\ 
19416 & 189.1381716 & 62.0901795  & $-99$        & $-99$  & 0.3834 & 0.38
& 0.3801 & 2650.6 & 172.872 & 100    & $-99$  \\ 
\tableline
\end{tabular}
\tablecomments{\raggedright This is a short version extracted from the released
  catalog. The complete table contains 63 columns for the 93598
  $zJHK_{{s}}$ detections. See detailed column description in
  Sec.~\ref{column_description} \\ }
\end{table*}

\section{Summary} \label{summary}

Homogeneous NIR images ($J$, $H$, and $K_{s}$) taken
by the instrument CFHT/WIRCam in the extended GOODS-N achieve
5$\sigma$ limiting AB magnitudes in 2\arcsec\ diameter apertures of
24.7, 24.2, and 24.4 mag in the $J$, $H$, and $K_{s}$ bands,
respectively. The images are released along with the multi-band
photometry and photometric redshift catalog, which contains $93598$
$zJHK_{{s}}$ detections.
\begin{enumerate}
    \item For the non-X-ray sources, we obtained accuracy
$\sigma_{\mathrm{NMAD}}=0.025$ and an outlier fraction $\eta=1.58\%$
for the $R<23$ sources. For the overall sample, we reached
$\sigma_{\mathrm{NMAD}}=0.036$ with $\eta=7.3\%$.
    \item For the X-ray sources, identified with the updated
X-ray catalog from \citet{xue16}, 
$\sigma_{\mathrm{NMAD}}=0.040$ with 
$\eta=10.5\%$. This outlier fraction is smaller than that of
previous work from the 3D-HST catalog.
    \item The use of UV data for the X-ray sample slightly improves the
{\phz} accuracy and reduces $\eta$ from
$9.8\%$ to $6.3\%$.
    
\end{enumerate}

These data  provide a large sample of objects to study
galaxy formation and evolution and the coevolution between AGNs
and their host galaxies. Future updates of the catalog will be
on the website: \url{http://idv.sinica.edu.tw/lthsu}.
Upcoming medium- or narrow-band data, e.g., SHARDS, will
improve {\phz} quality, especially for AGNs and star-forming
galaxies whose emission lines are usually strong.

\acknowledgments We are grateful to the referee for  constructive
comments. We thank Nicole Meger for her contributions on the analysis
of the Canadian $K$-band data. The work is supported by the Ministry
of Science \& Technology of Taiwan under  grants MOST
103-2112-M-001-031-MY3 and 106-2112-M-001-034. This work is based on observations
obtained with WIRCam, a joint project of CFHT, the Academia Sinica
Institute of Astronomy and Astrophysics (ASIAA) in Taiwan, the Korea
Astronomy and Space Science Institute (KASI) in Korea, and
the Canada-France-Hawaii Telescope (CFHT), which is
operated by the National Research Council (NRC) of Canada, the
Institut National des Sciences de l'Univers of the Centre National de
la Recherche Scientifique of France, and the University of Hawaii. 

Facilities: {\it GALEX}, {\it CHFT}/WIRCam, {\it Spitzer}/IRAC,
{Subaru}/Suprime-Cam, KPNO Mayall 4m telescope

\bibliography{literature_ecdfn}

\begin{thebibliography}{}
\expandafter\ifx\csname natexlab\endcsname\relax\def\natexlab#1{#1}\fi
\providecommand{\url}[1]{\href{#1}{#1}}
\providecommand{\dodoi}[1]{doi:~\href{http://doi.org/#1}{\nolinkurl{#1}}}
\providecommand{\doeprint}[1]{\href{http://ascl.net/#1}{\nolinkurl{http://ascl.net/#1}}}
\providecommand{\doarXiv}[1]{\href{https://arxiv.org/abs/#1}{\nolinkurl{https://arxiv.org/abs/#1}}}

\bibitem[{{Alexander} {et~al.}(2003){Alexander}, {Bauer}, {Brandt},
  {Schneider}, {Hornschemeier}, {Vignali}, {Barger}, {Broos}, {Cowie},
  {Garmire}, {Townsley}, {Bautz}, {Chartas}, \& {Sargent}}]{alexander03}
{Alexander}, D.~M., {Bauer}, F.~E., {Brandt}, W.~N., {et~al.} 2003, \aj, 126,
  539, \dodoi{10.1086/376473}

\bibitem[{{Arnouts} {et~al.}(1999){Arnouts}, {Cristiani}, {Moscardini},
  {Matarrese}, {Lucchin}, {Fontana}, \& {Giallongo}}]{arnouts99}
{Arnouts}, S., {Cristiani}, S., {Moscardini}, L., {et~al.} 1999, \mnras, 310,
  540, \dodoi{10.1046/j.1365-8711.1999.02978.x}

\bibitem[{{Ashby} {et~al.}(2013){Ashby}, {Willner}, {Fazio}, {Huang}, {Arendt},
  {Barmby}, {Barro}, {Bell}, {Bouwens}, {Cattaneo}, {Croton}, {Dav{\'e}},
  {Dunlop}, {Egami}, {Faber}, {Finlator}, {Grogin}, {Guhathakurta},
  {Hernquist}, {Hora}, {Illingworth}, {Kashlinsky}, {Koekemoer}, {Koo},
  {Labb{\'e}}, {Li}, {Lin}, {Moseley}, {Nandra}, {Newman}, {Noeske}, {Ouchi},
  {Peth}, {Rigopoulou}, {Robertson}, {Sarajedini}, {Simard}, {Smith}, {Wang},
  {Wechsler}, {Weiner}, {Wilson}, {Wuyts}, {Yamada}, \& {Yan}}]{ashby13}
{Ashby}, M.~L.~N., {Willner}, S.~P., {Fazio}, G.~G., {et~al.} 2013, \apj, 769,
  80, \dodoi{10.1088/0004-637X/769/1/80}

\bibitem[{{Ashby} {et~al.}(2015){Ashby}, {Willner}, {Fazio}, {Dunlop}, {Egami},
  {Faber}, {Ferguson}, {Grogin}, {Hora}, {Huang}, {Koekemoer}, {Labb{\'e}}, \&
  {Wang}}]{ashby2015}
---. 2015, \apjs, 218, 33, \dodoi{10.1088/0067-0049/218/2/33}

\bibitem[{{Barger} {et~al.}(2008){Barger}, {Cowie}, \& {Wang}}]{barger08}
{Barger}, A.~J., {Cowie}, L.~L., \& {Wang}, W.-H. 2008, \apj, 689, 687,
  \dodoi{10.1086/592735}

\bibitem[{{Bender} {et~al.}(2001){Bender}, {Appenzeller}, {B{\"o}hm}, {Drory},
  {Fricke}, {Gabasch}, {Heidt}, {Hopp}, {J{\"a}ger}, {K{\"u}mmel}, {Mehlert},
  {M{\"o}llenhoff}, {Moorwood}, {Nicklas}, {Noll}, {Saglia}, {Seifert},
  {Seitz}, {Stahl}, {Sutorius}, {Szeifert}, {Wagner}, \& {Ziegler}}]{bender01}
{Bender}, R., {Appenzeller}, I., {B{\"o}hm}, A., {et~al.} 2001, in Deep Fields,
  ed. S.~{Cristiani}, A.~{Renzini}, \& R.~E. {Williams}, 96

\bibitem[{{Bertin}(2006)}]{bertin06}
{Bertin}, E. 2006, in Astronomical Society of the Pacific Conference Series,
  Vol. 351, Astronomical Data Analysis Software and Systems XV, ed.
  C.~{Gabriel}, C.~{Arviset}, D.~{Ponz}, \& S.~{Enrique}, 112

\bibitem[{{Bertin} \& {Arnouts}(1996)}]{bertin96}
{Bertin}, E., \& {Arnouts}, S. 1996, \aaps, 117, 393,
  \dodoi{10.1051/aas:1996164}

\bibitem[{{Bertin} {et~al.}(2002){Bertin}, {Mellier}, {Radovich}, {Missonnier},
  {Didelon}, \& {Morin}}]{bertin02}
{Bertin}, E., {Mellier}, Y., {Radovich}, M., {et~al.} 2002, in Astronomical
  Society of the Pacific Conference Series, Vol. 281, Astronomical Data
  Analysis Software and Systems XI, ed. D.~A. {Bohlender}, D.~{Durand}, \&
  T.~H. {Handley}, 228

\bibitem[{{Bielby} {et~al.}(2012){Bielby}, {Hudelot}, {McCracken}, {Ilbert},
  {Daddi}, {Le F{\`e}vre}, {Gonzalez-Perez}, {Kneib}, {Marmo}, {Mellier},
  {Salvato}, {Sanders}, \& {Willott}}]{bielby2012}
{Bielby}, R., {Hudelot}, P., {McCracken}, H.~J., {et~al.} 2012, \aap, 545, A23,
  \dodoi{10.1051/0004-6361/201118547}

\bibitem[{{Bouwens} {et~al.}(2010){Bouwens}, {Illingworth}, {Gonz{\'a}lez},
  {Labb{\'e}}, {Franx}, {Conselice}, {Blakeslee}, {van Dokkum}, {Holden},
  {Magee}, {Marchesini}, \& {Zheng}}]{bouwens2010}
{Bouwens}, R.~J., {Illingworth}, G.~D., {Gonz{\'a}lez}, V., {et~al.} 2010,
  \apj, 725, 1587, \dodoi{10.1088/0004-637X/725/2/1587}

\bibitem[{{Bouwens} {et~al.}(2015){Bouwens}, {Illingworth}, {Oesch}, {Trenti},
  {Labb{\'e}}, {Bradley}, {Carollo}, {van Dokkum}, {Gonzalez}, {Holwerda},
  {Franx}, {Spitler}, {Smit}, \& {Magee}}]{bouwens2015}
{Bouwens}, R.~J., {Illingworth}, G.~D., {Oesch}, P.~A., {et~al.} 2015, \apj,
  803, 34, \dodoi{10.1088/0004-637X/803/1/34}

\bibitem[{{Brammer} {et~al.}(2008){Brammer}, {van Dokkum}, \&
  {Coppi}}]{brammer2008}
{Brammer}, G.~B., {van Dokkum}, P.~G., \& {Coppi}, P. 2008, \apj, 686, 1503,
  \dodoi{10.1086/591786}

\bibitem[{{Bruzual} \& {Charlot}(2003)}]{BC03}
{Bruzual}, G., \& {Charlot}, S. 2003, \mnras, 344, 1000,
  \dodoi{10.1046/j.1365-8711.2003.06897.x}

\bibitem[{{Calzetti} {et~al.}(2000){Calzetti}, {Armus}, {Bohlin}, {Kinney},
  {Koornneef}, \& {Storchi-Bergmann}}]{calzetti00}
{Calzetti}, D., {Armus}, L., {Bohlin}, R.~C., {et~al.} 2000, \apj, 533, 682,
  \dodoi{10.1086/308692}

\bibitem[{{Capak} {et~al.}(2004){Capak}, {Cowie}, {Hu}, {Barger}, {Dickinson},
  {Fernandez}, {Giavalisco}, {Komiyama}, {Kretchmer}, {McNally}, {Miyazaki},
  {Okamura}, \& {Stern}}]{capak04}
{Capak}, P., {Cowie}, L.~L., {Hu}, E.~M., {et~al.} 2004, \aj, 127, 180,
  \dodoi{10.1086/380611}

\bibitem[{{Capak} {et~al.}(2007){Capak}, {Aussel}, {Ajiki}, {McCracken},
  {Mobasher}, {Scoville}, {Shopbell}, {Taniguchi}, {Thompson}, {Tribiano},
  {Sasaki}, {Blain}, {Brusa}, {Carilli}, {Comastri}, {Carollo}, {Cassata},
  {Colbert}, {Ellis}, {Elvis}, {Giavalisco}, {Green}, {Guzzo}, {Hasinger},
  {Ilbert}, {Impey}, {Jahnke}, {Kartaltepe}, {Kneib}, {Koda}, {Koekemoer},
  {Komiyama}, {Leauthaud}, {Le Fevre}, {Lilly}, {Liu}, {Massey}, {Miyazaki},
  {Murayama}, {Nagao}, {Peacock}, {Pickles}, {Porciani}, {Renzini}, {Rhodes},
  {Rich}, {Salvato}, {Sanders}, {Scarlata}, {Schiminovich}, {Schinnerer},
  {Scodeggio}, {Sheth}, {Shioya}, {Tasca}, {Taylor}, {Yan}, \&
  {Zamorani}}]{capak2007}
{Capak}, P., {Aussel}, H., {Ajiki}, M., {et~al.} 2007, \apjs, 172, 99,
  \dodoi{10.1086/519081}

\bibitem[{{Cassata} {et~al.}(2011){Cassata}, {Giavalisco}, {Guo}, {Renzini},
  {Ferguson}, {Koekemoer}, {Salimbeni}, {Scarlata}, {Grogin}, {Conselice},
  {Dahlen}, {Lotz}, {Dickinson}, \& {Lin}}]{cassata11}
{Cassata}, P., {Giavalisco}, M., {Guo}, Y., {et~al.} 2011, \apj, 743, 96,
  \dodoi{10.1088/0004-637X/743/1/96}

\bibitem[{{Cooper} {et~al.}(2011){Cooper}, {Aird}, {Coil}, {Davis}, {Faber},
  {Juneau}, {Lotz}, {Nandra}, {Newman}, {Willmer}, \& {Yan}}]{cooper2011}
{Cooper}, M.~C., {Aird}, J.~A., {Coil}, A.~L., {et~al.} 2011, \apjs, 193, 14,
  \dodoi{10.1088/0067-0049/193/1/14}

\bibitem[{{Cowie} {et~al.}(2004){Cowie}, {Barger}, {Hu}, {Capak}, \&
  {Songaila}}]{cowie04}
{Cowie}, L.~L., {Barger}, A.~J., {Hu}, E.~M., {Capak}, P., \& {Songaila}, A.
  2004, \aj, 127, 3137, \dodoi{10.1086/420997}

\bibitem[{{Daddi} {et~al.}(2004){Daddi}, {Cimatti}, {Renzini}, {Fontana},
  {Mignoli}, {Pozzetti}, {Tozzi}, \& {Zamorani}}]{daddi04}
{Daddi}, E., {Cimatti}, A., {Renzini}, A., {et~al.} 2004, \apj, 617, 746,
  \dodoi{10.1086/425569}

\bibitem[{{Dahlen} {et~al.}(2010){Dahlen}, {Mobasher}, {Dickinson}, {Ferguson},
  {Giavalisco}, {Grogin}, {Guo}, {Koekemoer}, {Lee}, {Lee}, {Nonino}, {Riess},
  \& {Salimbeni}}]{dahlen2010}
{Dahlen}, T., {Mobasher}, B., {Dickinson}, M., {et~al.} 2010, \apj, 724, 425,
  \dodoi{10.1088/0004-637X/724/1/425}

\bibitem[{{Dey} {et~al.}(2008){Dey}, {Soifer}, {Desai}, {Brand}, {Le Floc'h},
  {Brown}, {Jannuzi}, {Armus}, {Bussmann}, {Brodwin}, {Bian}, {Eisenhardt},
  {Higdon}, {Weedman}, \& {Willner}}]{dey08}
{Dey}, A., {Soifer}, B.~T., {Desai}, V., {et~al.} 2008, \apj, 677, 943,
  \dodoi{10.1086/529516}

\bibitem[{{Elbaz} {et~al.}(2011){Elbaz}, {Dickinson}, {Hwang},
  {D{\'{\i}}az-Santos}, {Magdis}, {Magnelli}, {Le Borgne}, {Galliano},
  {Pannella}, {Chanial}, {Armus}, {Charmandaris}, {Daddi}, {Aussel}, {Popesso},
  {Kartaltepe}, {Altieri}, {Valtchanov}, {Coia}, {Dannerbauer}, {Dasyra},
  {Leiton}, {Mazzarella}, {Alexander}, {Buat}, {Burgarella}, {Chary}, {Gilli},
  {Ivison}, {Juneau}, {Le Floc'h}, {Lutz}, {Morrison}, {Mullaney}, {Murphy},
  {Pope}, {Scott}, {Brodwin}, {Calzetti}, {Cesarsky}, {Charlot}, {Dole},
  {Eisenhardt}, {Ferguson}, {F{\"o}rster Schreiber}, {Frayer}, {Giavalisco},
  {Huynh}, {Koekemoer}, {Papovich}, {Reddy}, {Surace}, {Teplitz}, {Yun}, \&
  {Wilson}}]{elbaz11}
{Elbaz}, D., {Dickinson}, M., {Hwang}, H.~S., {et~al.} 2011, \aap, 533, A119,
  \dodoi{10.1051/0004-6361/201117239}

\bibitem[{{Elston} {et~al.}(1988){Elston}, {Rieke}, \& {Rieke}}]{elston88}
{Elston}, R., {Rieke}, G.~H., \& {Rieke}, M.~J. 1988, \apjl, 331, L77,
  \dodoi{10.1086/185239}

\bibitem[{{Franx} {et~al.}(2003){Franx}, {Labb{\'e}}, {Rudnick}, {van Dokkum},
  {Daddi}, {F{\"o}rster Schreiber}, {Moorwood}, {Rix}, {R{\"o}ttgering}, {van
  der Wel}, {van der Werf}, \& {van Starkenburg}}]{franx03}
{Franx}, M., {Labb{\'e}}, I., {Rudnick}, G., {et~al.} 2003, \apjl, 587, L79,
  \dodoi{10.1086/375155}

\bibitem[{{Giavalisco}(2012)}]{giavalisco2012}
{Giavalisco}, M. 2012, VizieR Online Data Catalog, 2318

\bibitem[{{Giavalisco} {et~al.}(2004){Giavalisco}, {Ferguson}, {Koekemoer},
  {Dickinson}, {Alexander}, {Bauer}, {Bergeron}, {Biagetti}, {Brandt},
  {Casertano}, {Cesarsky}, {Chatzichristou}, {Conselice}, {Cristiani}, {Da
  Costa}, {Dahlen}, {de Mello}, {Eisenhardt}, {Erben}, {Fall}, {Fassnacht},
  {Fosbury}, {Fruchter}, {Gardner}, {Grogin}, {Hook}, {Hornschemeier}, {Idzi},
  {Jogee}, {Kretchmer}, {Laidler}, {Lee}, {Livio}, {Lucas}, {Madau},
  {Mobasher}, {Moustakas}, {Nonino}, {Padovani}, {Papovich}, {Park},
  {Ravindranath}, {Renzini}, {Richardson}, {Riess}, {Rosati}, {Schirmer},
  {Schreier}, {Somerville}, {Spinrad}, {Stern}, {Stiavelli}, {Strolger},
  {Urry}, {Vandame}, {Williams}, \& {Wolf}}]{giavalisco04}
{Giavalisco}, M., {Ferguson}, H.~C., {Koekemoer}, A.~M., {et~al.} 2004, \apjl,
  600, L93, \dodoi{10.1086/379232}

\bibitem[{{Grogin} {et~al.}(2011){Grogin}, {Kocevski}, {Faber}, {Ferguson},
  {Koekemoer}, {Riess}, {Acquaviva}, {Alexander}, {Almaini}, {Ashby}, {Barden},
  {Bell}, {Bournaud}, {Brown}, {Caputi}, {Casertano}, {Cassata}, {Castellano},
  {Challis}, {Chary}, {Cheung}, {Cirasuolo}, {Conselice}, {Roshan Cooray},
  {Croton}, {Daddi}, {Dahlen}, {Dav{\'e}}, {de Mello}, {Dekel}, {Dickinson},
  {Dolch}, {Donley}, {Dunlop}, {Dutton}, {Elbaz}, {Fazio}, {Filippenko},
  {Finkelstein}, {Fontana}, {Gardner}, {Garnavich}, {Gawiser}, {Giavalisco},
  {Grazian}, {Guo}, {Hathi}, {H{\"a}ussler}, {Hopkins}, {Huang}, {Huang},
  {Jha}, {Kartaltepe}, {Kirshner}, {Koo}, {Lai}, {Lee}, {Li}, {Lotz}, {Lucas},
  {Madau}, {McCarthy}, {McGrath}, {McIntosh}, {McLure}, {Mobasher},
  {Moustakas}, {Mozena}, {Nandra}, {Newman}, {Niemi}, {Noeske}, {Papovich},
  {Pentericci}, {Pope}, {Primack}, {Rajan}, {Ravindranath}, {Reddy}, {Renzini},
  {Rix}, {Robaina}, {Rodney}, {Rosario}, {Rosati}, {Salimbeni}, {Scarlata},
  {Siana}, {Simard}, {Smidt}, {Somerville}, {Spinrad}, {Straughn}, {Strolger},
  {Telford}, {Teplitz}, {Trump}, {van der Wel}, {Villforth}, {Wechsler},
  {Weiner}, {Wiklind}, {Wild}, {Wilson}, {Wuyts}, {Yan}, \& {Yun}}]{grogin11}
{Grogin}, N.~A., {Kocevski}, D.~D., {Faber}, S.~M., {et~al.} 2011, \apjs, 197,
  35, \dodoi{10.1088/0067-0049/197/2/35}

\bibitem[{{Guo} {et~al.}(2012){Guo}, {Giavalisco}, {Cassata}, {Ferguson},
  {Williams}, {Dickinson}, {Koekemoer}, {Grogin}, {Chary}, {Messias}, {Tundo},
  {Lin}, {Lee}, {Salimbeni}, {Fontana}, {Grazian}, {Kocevski}, {Lee},
  {Villanueva}, \& {van der Wel}}]{guo12}
{Guo}, Y., {Giavalisco}, M., {Cassata}, P., {et~al.} 2012, \apj, 749, 149,
  \dodoi{10.1088/0004-637X/749/2/149}

\bibitem[{{Hsieh} {et~al.}(2012){Hsieh}, {Wang}, {Hsieh}, {Lin}, {Yan}, {Lim},
  \& {Ho}}]{hsieh12}
{Hsieh}, B.-C., {Wang}, W.-H., {Hsieh}, C.-C., {et~al.} 2012, \apjs, 203, 23,
  \dodoi{10.1088/0067-0049/203/2/23}

\bibitem[{{Hsu} {et~al.}(2014){Hsu}, {Salvato}, {Nandra}, {Brusa}, {Bender},
  {Buchner}, {Donley}, {Kocevski}, {Guo}, {Hathi}, {Rangel}, {Willner},
  {Brightman}, {Georgakakis}, {Budav{\'a}ri}, {Szalay}, {Ashby}, {Barro},
  {Dahlen}, {Faber}, {Ferguson}, {Galametz}, {Grazian}, {Grogin}, {Huang},
  {Koekemoer}, {Lucas}, {McGrath}, {Mobasher}, {Peth}, {Rosario}, \&
  {Trump}}]{hsu14}
{Hsu}, L.-T., {Salvato}, M., {Nandra}, K., {et~al.} 2014, \apj, 796, 60,
  \dodoi{10.1088/0004-637X/796/1/60}

\bibitem[{{Ilbert} {et~al.}(2006){Ilbert}, {Arnouts}, {McCracken},
  {Bolzonella}, {Bertin}, {Le F{\`e}vre}, {Mellier}, {Zamorani}, {Pell{\`o}},
  {Iovino}, {Tresse}, {Le Brun}, {Bottini}, {Garilli}, {Maccagni}, {Picat},
  {Scaramella}, {Scodeggio}, {Vettolani}, {Zanichelli}, {Adami}, {Bardelli},
  {Cappi}, {Charlot}, {Ciliegi}, {Contini}, {Cucciati}, {Foucaud}, {Franzetti},
  {Gavignaud}, {Guzzo}, {Marano}, {Marinoni}, {Mazure}, {Meneux}, {Merighi},
  {Paltani}, {Pollo}, {Pozzetti}, {Radovich}, {Zucca}, {Bondi}, {Bongiorno},
  {Busarello}, {de La Torre}, {Gregorini}, {Lamareille}, {Mathez}, {Merluzzi},
  {Ripepi}, {Rizzo}, \& {Vergani}}]{ilbert06}
{Ilbert}, O., {Arnouts}, S., {McCracken}, H.~J., {et~al.} 2006, \aap, 457, 841,
  \dodoi{10.1051/0004-6361:20065138}

\bibitem[{{Ilbert} {et~al.}(2009){Ilbert}, {Capak}, {Salvato}, {Aussel},
  {McCracken}, {Sanders}, {Scoville}, {Kartaltepe}, {Arnouts}, {Le Floc'h},
  {Mobasher}, {Taniguchi}, {Lamareille}, {Leauthaud}, {Sasaki}, {Thompson},
  {Zamojski}, {Zamorani}, {Bardelli}, {Bolzonella}, {Bongiorno}, {Brusa},
  {Caputi}, {Carollo}, {Contini}, {Cook}, {Coppa}, {Cucciati}, {de la Torre},
  {de Ravel}, {Franzetti}, {Garilli}, {Hasinger}, {Iovino}, {Kampczyk},
  {Kneib}, {Knobel}, {Kovac}, {Le Borgne}, {Le Brun}, {F{\`e}vre}, {Lilly},
  {Looper}, {Maier}, {Mainieri}, {Mellier}, {Mignoli}, {Murayama}, {Pell{\`o}},
  {Peng}, {P{\'e}rez-Montero}, {Renzini}, {Ricciardelli}, {Schiminovich},
  {Scodeggio}, {Shioya}, {Silverman}, {Surace}, {Tanaka}, {Tasca}, {Tresse},
  {Vergani}, \& {Zucca}}]{ilbert09}
{Ilbert}, O., {Capak}, P., {Salvato}, M., {et~al.} 2009, \apj, 690, 1236,
  \dodoi{10.1088/0004-637X/690/2/1236}

\bibitem[{{Ilbert} {et~al.}(2013){Ilbert}, {McCracken}, {Le F{\`e}vre},
  {Capak}, {Dunlop}, {Karim}, {Renzini}, {Caputi}, {Boissier}, {Arnouts},
  {Aussel}, {Comparat}, {Guo}, {Hudelot}, {Kartaltepe}, {Kneib}, {Krogager},
  {Le Floc'h}, {Lilly}, {Mellier}, {Milvang-Jensen}, {Moutard}, {Onodera},
  {Richard}, {Salvato}, {Sanders}, {Scoville}, {Silverman}, {Taniguchi},
  {Tasca}, {Thomas}, {Toft}, {Tresse}, {Vergani}, {Wolk}, \&
  {Zirm}}]{ilbert2013}
{Ilbert}, O., {McCracken}, H.~J., {Le F{\`e}vre}, O., {et~al.} 2013, \aap, 556,
  A55, \dodoi{10.1051/0004-6361/201321100}

\bibitem[{{Kajisawa} {et~al.}(2011){Kajisawa}, {Ichikawa}, {Tanaka}, {Yamada},
  {Akiyama}, {Suzuki}, {Tokoku}, {Katsuno Uchimoto}, {Konishi}, {Yoshikawa},
  {Nishimura}, {Omata}, {Ouchi}, {Iwata}, {Hamana}, \& {Onodera}}]{kajisawa11}
{Kajisawa}, M., {Ichikawa}, T., {Tanaka}, I., {et~al.} 2011, \pasj, 63, 379,
  \dodoi{10.1093/pasj/63.sp2.S379}

\bibitem[{{Keenan} {et~al.}(2010){Keenan}, {Trouille}, {Barger}, {Cowie}, \&
  {Wang}}]{keenan10}
{Keenan}, R.~C., {Trouille}, L., {Barger}, A.~J., {Cowie}, L.~L., \& {Wang},
  W.-H. 2010, \apjs, 186, 94, \dodoi{10.1088/0067-0049/186/1/94}

\bibitem[{{Kennicutt}(1998)}]{kennicutt98}
{Kennicutt}, Jr., R.~C. 1998, \araa, 36, 189,
  \dodoi{10.1146/annurev.astro.36.1.189}

\bibitem[{{Koekemoer} {et~al.}(2011){Koekemoer}, {Faber}, {Ferguson}, {Grogin},
  {Kocevski}, {Koo}, {Lai}, {Lotz}, {Lucas}, {McGrath}, {Ogaz}, {Rajan},
  {Riess}, {Rodney}, {Strolger}, {Casertano}, {Castellano}, {Dahlen},
  {Dickinson}, {Dolch}, {Fontana}, {Giavalisco}, {Grazian}, {Guo}, {Hathi},
  {Huang}, {van der Wel}, {Yan}, {Acquaviva}, {Alexander}, {Almaini}, {Ashby},
  {Barden}, {Bell}, {Bournaud}, {Brown}, {Caputi}, {Cassata}, {Challis},
  {Chary}, {Cheung}, {Cirasuolo}, {Conselice}, {Roshan Cooray}, {Croton},
  {Daddi}, {Dav{\'e}}, {de Mello}, {de Ravel}, {Dekel}, {Donley}, {Dunlop},
  {Dutton}, {Elbaz}, {Fazio}, {Filippenko}, {Finkelstein}, {Frazer}, {Gardner},
  {Garnavich}, {Gawiser}, {Gruetzbauch}, {Hartley}, {H{\"a}ussler},
  {Herrington}, {Hopkins}, {Huang}, {Jha}, {Johnson}, {Kartaltepe},
  {Khostovan}, {Kirshner}, {Lani}, {Lee}, {Li}, {Madau}, {McCarthy},
  {McIntosh}, {McLure}, {McPartland}, {Mobasher}, {Moreira}, {Mortlock},
  {Moustakas}, {Mozena}, {Nandra}, {Newman}, {Nielsen}, {Niemi}, {Noeske},
  {Papovich}, {Pentericci}, {Pope}, {Primack}, {Ravindranath}, {Reddy},
  {Renzini}, {Rix}, {Robaina}, {Rosario}, {Rosati}, {Salimbeni}, {Scarlata},
  {Siana}, {Simard}, {Smidt}, {Snyder}, {Somerville}, {Spinrad}, {Straughn},
  {Telford}, {Teplitz}, {Trump}, {Vargas}, {Villforth}, {Wagner}, {Wandro},
  {Wechsler}, {Weiner}, {Wiklind}, {Wild}, {Wilson}, {Wuyts}, \&
  {Yun}}]{koekemoer11}
{Koekemoer}, A.~M., {Faber}, S.~M., {Ferguson}, H.~C., {et~al.} 2011, \apjs,
  197, 36, \dodoi{10.1088/0067-0049/197/2/36}

\bibitem[{{Kron}(1980)}]{kron1980}
{Kron}, R.~G. 1980, \apjs, 43, 305, \dodoi{10.1086/190669}

\bibitem[{{Laigle} {et~al.}(2016){Laigle}, {McCracken}, {Ilbert}, {Hsieh},
  {Davidzon}, {Capak}, {Hasinger}, {Silverman}, {Pichon}, {Coupon}, {Aussel},
  {Le Borgne}, {Caputi}, {Cassata}, {Chang}, {Civano}, {Dunlop}, {Fynbo},
  {Kartaltepe}, {Koekemoer}, {Le F{\`e}vre}, {Le Floc'h}, {Leauthaud}, {Lilly},
  {Lin}, {Marchesi}, {Milvang-Jensen}, {Salvato}, {Sanders}, {Scoville},
  {Smolcic}, {Stockmann}, {Taniguchi}, {Tasca}, {Toft}, {Vaccari}, \&
  {Zabl}}]{laigle16}
{Laigle}, C., {McCracken}, H.~J., {Ilbert}, O., {et~al.} 2016, \apjs, 224, 24,
  \dodoi{10.3847/0067-0049/224/2/24}

\bibitem[{{Lawrence} {et~al.}(2007){Lawrence}, {Warren}, {Almaini}, {Edge},
  {Hambly}, {Jameson}, {Lucas}, {Casali}, {Adamson}, {Dye}, {Emerson},
  {Foucaud}, {Hewett}, {Hirst}, {Hodgkin}, {Irwin}, {Lodieu}, {McMahon},
  {Simpson}, {Smail}, {Mortlock}, \& {Folger}}]{lawrence07}
{Lawrence}, A., {Warren}, S.~J., {Almaini}, O., {et~al.} 2007, \mnras, 379,
  1599, \dodoi{10.1111/j.1365-2966.2007.12040.x}

\bibitem[{{Lee} {et~al.}(2012){Lee}, {Ferguson}, {Wiklind}, {Dahlen},
  {Dickinson}, {Giavalisco}, {Grogin}, {Papovich}, {Messias}, {Guo}, \&
  {Lin}}]{lee12}
{Lee}, K.-S., {Ferguson}, H.~C., {Wiklind}, T., {et~al.} 2012, \apj, 752, 66,
  \dodoi{10.1088/0004-637X/752/1/66}

\bibitem[{{Lin} {et~al.}(2012){Lin}, {Dickinson}, {Jian}, {Merson}, {Baugh},
  {Scott}, {Foucaud}, {Wang}, {Yan}, {Yan}, {Cheng}, {Guo}, {Helly}, {Kirsten},
  {Koo}, {Lagos}, {Meger}, {Messias}, {Pope}, {Simard}, {Grogin}, \&
  {Wang}}]{lin12}
{Lin}, L., {Dickinson}, M., {Jian}, H.-Y., {et~al.} 2012, \apj, 756, 71,
  \dodoi{10.1088/0004-637X/756/1/71}

\bibitem[{{McCracken} {et~al.}(2012){McCracken}, {Milvang-Jensen}, {Dunlop},
  {Franx}, {Fynbo}, {Le F{\`e}vre}, {Holt}, {Caputi}, {Goranova}, {Buitrago},
  {Emerson}, {Freudling}, {Hudelot}, {L{\'o}pez-Sanjuan}, {Magnard}, {Mellier},
  {M{\o}ller}, {Nilsson}, {Sutherland}, {Tasca}, \& {Zabl}}]{McCracken12}
{McCracken}, H.~J., {Milvang-Jensen}, B., {Dunlop}, J., {et~al.} 2012, \aap,
  544, A156, \dodoi{10.1051/0004-6361/201219507}

\bibitem[{{Morrison} {et~al.}(2010){Morrison}, {Owen}, {Dickinson}, {Ivison},
  \& {Ibar}}]{morrison10}
{Morrison}, G.~E., {Owen}, F.~N., {Dickinson}, M., {Ivison}, R.~J., \& {Ibar},
  E. 2010, \apjs, 188, 178, \dodoi{10.1088/0067-0049/188/1/178}

\bibitem[{{Murphy} {et~al.}(2011){Murphy}, {Chary}, {Dickinson}, {Pope},
  {Frayer}, \& {Lin}}]{murphy11}
{Murphy}, E.~J., {Chary}, R.-R., {Dickinson}, M., {et~al.} 2011, \apj, 732,
  126, \dodoi{10.1088/0004-637X/732/2/126}

\bibitem[{{Onodera} {et~al.}(2012){Onodera}, {Renzini}, {Carollo},
  {Cappellari}, {Mancini}, {Strazzullo}, {Daddi}, {Arimoto}, {Gobat}, {Yamada},
  {McCracken}, {Ilbert}, {Capak}, {Cimatti}, {Giavalisco}, {Koekemoer}, {Kong},
  {Lilly}, {Motohara}, {Ohta}, {Sanders}, {Scoville}, {Tamura}, \&
  {Taniguchi}}]{onodera2012}
{Onodera}, M., {Renzini}, A., {Carollo}, M., {et~al.} 2012, \apj, 755, 26,
  \dodoi{10.1088/0004-637X/755/1/26}

\bibitem[{{Ouchi} {et~al.}(2009){Ouchi}, {Mobasher}, {Shimasaku}, {Ferguson},
  {Fall}, {Ono}, {Kashikawa}, {Morokuma}, {Nakajima}, {Okamura}, {Dickinson},
  {Giavalisco}, \& {Ohta}}]{ouchi09}
{Ouchi}, M., {Mobasher}, B., {Shimasaku}, K., {et~al.} 2009, \apj, 706, 1136,
  \dodoi{10.1088/0004-637X/706/2/1136}

\bibitem[{{Penner} {et~al.}(2012){Penner}, {Dickinson}, {Pope}, {Dey},
  {Magnelli}, {Pannella}, {Altieri}, {Aussel}, {Buat}, {Bussmann},
  {Charmandaris}, {Coia}, {Daddi}, {Dannerbauer}, {Elbaz}, {Hwang},
  {Kartaltepe}, {Lin}, {Magdis}, {Morrison}, {Popesso}, {Scott}, \&
  {Valtchanov}}]{penner12}
{Penner}, K., {Dickinson}, M., {Pope}, A., {et~al.} 2012, \apj, 759, 28,
  \dodoi{10.1088/0004-637X/759/1/28}

\bibitem[{{P{\'e}rez-Gonz{\'a}lez} \& {Cava}(2013)}]{shards2013}
{P{\'e}rez-Gonz{\'a}lez}, P.~G., \& {Cava}, A. 2013, in Revista Mexicana de
  Astronomia y Astrofisica Conference Series, Vol.~42, Revista Mexicana de
  Astronomia y Astrofisica Conference Series, 55--57

\bibitem[{{Polletta} {et~al.}(2007){Polletta}, {Tajer}, {Maraschi},
  {Trinchieri}, {Lonsdale}, {Chiappetti}, {Andreon}, {Pierre}, {Le F{\`e}vre},
  {Zamorani}, {Maccagni}, {Garcet}, {Surdej}, {Franceschini}, {Alloin},
  {Shupe}, {Surace}, {Fang}, {Rowan-Robinson}, {Smith}, \&
  {Tresse}}]{polletta07}
{Polletta}, M., {Tajer}, M., {Maraschi}, L., {et~al.} 2007, \apj, 663, 81,
  \dodoi{10.1086/518113}

\bibitem[{{Prevot} {et~al.}(1984){Prevot}, {Lequeux}, {Prevot}, {Maurice}, \&
  {Rocca-Volmerange}}]{prevot84}
{Prevot}, M.~L., {Lequeux}, J., {Prevot}, L., {Maurice}, E., \&
  {Rocca-Volmerange}, B. 1984, \aap, 132, 389

\bibitem[{{Puget} {et~al.}(2004){Puget}, {Stadler}, {Doyon}, {Gigan},
  {Thibault}, {Luppino}, {Barrick}, {Benedict}, {Forveille}, {Rambold},
  {Thomas}, {Vermeulen}, {Ward}, {Beuzit}, {Feautrier}, {Magnard}, {Mella},
  {Preis}, {Vallee}, {Wang}, {Lin}, {Hall}, \& {Hodapp}}]{puget04}
{Puget}, P., {Stadler}, E., {Doyon}, R., {et~al.} 2004, in \procspie, Vol.
  5492, Ground-based Instrumentation for Astronomy, ed. A.~F.~M. {Moorwood} \&
  M.~{Iye}, 978--987

\bibitem[{{Rafferty} {et~al.}(2011){Rafferty}, {Brandt}, {Alexander}, {Xue},
  {Bauer}, {Lehmer}, {Luo}, \& {Papovich}}]{rafferty11}
{Rafferty}, D.~A., {Brandt}, W.~N., {Alexander}, D.~M., {et~al.} 2011, \apj,
  742, 3, \dodoi{10.1088/0004-637X/742/1/3}

\bibitem[{{Salvato} {et~al.}(2009){Salvato}, {Hasinger}, {Ilbert}, {Zamorani},
  {Brusa}, {Scoville}, {Rau}, {Capak}, {Arnouts}, {Aussel}, {Bolzonella},
  {Buongiorno}, {Cappelluti}, {Caputi}, {Civano}, {Cook}, {Elvis}, {Gilli},
  {Jahnke}, {Kartaltepe}, {Impey}, {Lamareille}, {Le Floc'h}, {Lilly},
  {Mainieri}, {McCarthy}, {McCracken}, {Mignoli}, {Mobasher}, {Murayama},
  {Sasaki}, {Sanders}, {Schiminovich}, {Shioya}, {Shopbell}, {Silverman},
  {Smol{\v c}i{\'c}}, {Surace}, {Taniguchi}, {Thompson}, {Trump}, {Urry}, \&
  {Zamojski}}]{salvato09}
{Salvato}, M., {Hasinger}, G., {Ilbert}, O., {et~al.} 2009, \apj, 690, 1250,
  \dodoi{10.1088/0004-637X/690/2/1250}

\bibitem[{{Salvato} {et~al.}(2011){Salvato}, {Ilbert}, {Hasinger}, {Rau},
  {Civano}, {Zamorani}, {Brusa}, {Elvis}, {Vignali}, {Aussel}, {Comastri},
  {Fiore}, {Le Floc'h}, {Mainieri}, {Bardelli}, {Bolzonella}, {Bongiorno},
  {Capak}, {Caputi}, {Cappelluti}, {Carollo}, {Contini}, {Garilli}, {Iovino},
  {Fotopoulou}, {Fruscione}, {Gilli}, {Halliday}, {Kneib}, {Kakazu},
  {Kartaltepe}, {Koekemoer}, {Kovac}, {Ideue}, {Ikeda}, {Impey}, {Le Fevre},
  {Lamareille}, {Lanzuisi}, {Le Borgne}, {Le Brun}, {Lilly}, {Maier},
  {Manohar}, {Masters}, {McCracken}, {Messias}, {Mignoli}, {Mobasher}, {Nagao},
  {Pello}, {Puccetti}, {Perez-Montero}, {Renzini}, {Sargent}, {Sanders},
  {Scodeggio}, {Scoville}, {Shopbell}, {Silvermann}, {Taniguchi}, {Tasca},
  {Tresse}, {Trump}, \& {Zucca}}]{salvato11}
{Salvato}, M., {Ilbert}, O., {Hasinger}, G., {et~al.} 2011, \apj, 742, 61,
  \dodoi{10.1088/0004-637X/742/2/61}

\bibitem[{{Shim} {et~al.}(2011){Shim}, {Chary}, {Dickinson}, {Lin}, {Spinrad},
  {Stern}, \& {Yan}}]{shim11}
{Shim}, H., {Chary}, R.-R., {Dickinson}, M., {et~al.} 2011, \apj, 738, 69,
  \dodoi{10.1088/0004-637X/738/1/69}

\bibitem[{{Skelton} {et~al.}(2014){Skelton}, {Whitaker}, {Momcheva}, {Brammer},
  {van Dokkum}, {Labb{\'e}}, {Franx}, {van der Wel}, {Bezanson}, {Da Cunha},
  {Fumagalli}, {F{\"o}rster Schreiber}, {Kriek}, {Leja}, {Lundgren}, {Magee},
  {Marchesini}, {Maseda}, {Nelson}, {Oesch}, {Pacifici}, {Patel}, {Price},
  {Rix}, {Tal}, {Wake}, \& {Wuyts}}]{skelton14}
{Skelton}, R.~E., {Whitaker}, K.~E., {Momcheva}, I.~G., {et~al.} 2014, \apjs,
  214, 24, \dodoi{10.1088/0067-0049/214/2/24}

\bibitem[{{Skrutskie} {et~al.}(2006){Skrutskie}, {Cutri}, {Stiening},
  {Weinberg}, {Schneider}, {Carpenter}, {Beichman}, {Capps}, {Chester},
  {Elias}, {Huchra}, {Liebert}, {Lonsdale}, {Monet}, {Price}, {Seitzer},
  {Jarrett}, {Kirkpatrick}, {Gizis}, {Howard}, {Evans}, {Fowler}, {Fullmer},
  {Hurt}, {Light}, {Kopan}, {Marsh}, {McCallon}, {Tam}, {Van Dyk}, \&
  {Wheelock}}]{skrutskie06}
{Skrutskie}, M.~F., {Cutri}, R.~M., {Stiening}, R., {et~al.} 2006, \aj, 131,
  1163, \dodoi{10.1086/498708}

\bibitem[{{Steidel} {et~al.}(2003){Steidel}, {Adelberger}, {Shapley},
  {Pettini}, {Dickinson}, \& {Giavalisco}}]{steidel03}
{Steidel}, C.~C., {Adelberger}, K.~L., {Shapley}, A.~E., {et~al.} 2003, \apj,
  592, 728, \dodoi{10.1086/375772}

\bibitem[{{Teplitz} {et~al.}(2011){Teplitz}, {Chary}, {Elbaz}, {Dickinson},
  {Bridge}, {Colbert}, {Le Floc'h}, {Frayer}, {Howell}, {Koo}, {Papovich},
  {Phillips}, {Scarlata}, {Siana}, {Spinrad}, \& {Stern}}]{teplitz2011}
{Teplitz}, H.~I., {Chary}, R., {Elbaz}, D., {et~al.} 2011, \aj, 141, 1,
  \dodoi{10.1088/0004-6256/141/1/1}

\bibitem[{{Treister} {et~al.}(2006){Treister}, {Urry}, {Van Duyne},
  {Dickinson}, {Chary}, {Alexander}, {Bauer}, {Natarajan}, {Lira}, \&
  {Grogin}}]{treister2006}
{Treister}, E., {Urry}, C.~M., {Van Duyne}, J., {et~al.} 2006, \apj, 640, 603,
  \dodoi{10.1086/500237}

\bibitem[{{Wang}(2010)}]{wang10b}
{Wang}, W.-H. 2010, in Astronomical Society of the Pacific Conference Series,
  Vol. 434, Astronomical Data Analysis Software and Systems XIX, ed.
  Y.~{Mizumoto}, K.-I. {Morita}, \& M.~{Ohishi}, 87

\bibitem[{{Wang} {et~al.}(2010){Wang}, {Cowie}, {Barger}, {Keenan}, \&
  {Ting}}]{wang10a}
{Wang}, W.-H., {Cowie}, L.~L., {Barger}, A.~J., {Keenan}, R.~C., \& {Ting},
  H.-C. 2010, \apjs, 187, 251, \dodoi{10.1088/0067-0049/187/1/251}

\bibitem[{{Xue} {et~al.}(2016){Xue}, {Luo}, {Brandt}, {Alexander}, {Bauer},
  {Lehmer}, \& {Yang}}]{xue16}
{Xue}, Y.~Q., {Luo}, B., {Brandt}, W.~N., {et~al.} 2016, \apjs, 224, 15,
  \dodoi{10.3847/0067-0049/224/2/15}

\bibitem[{{Xue} {et~al.}(2011){Xue}, {Luo}, {Brandt}, {Bauer}, {Lehmer},
  {Broos}, {Schneider}, {Alexander}, {Brusa}, {Comastri}, {Fabian}, {Gilli},
  {Hasinger}, {Hornschemeier}, {Koekemoer}, {Liu}, {Mainieri}, {Paolillo},
  {Rafferty}, {Rosati}, {Shemmer}, {Silverman}, {Smail}, {Tozzi}, \&
  {Vignali}}]{xue11}
---. 2011, \apjs, 195, 10, \dodoi{10.1088/0067-0049/195/1/10}

\bibitem[{{Yang} {et~al.}(2014){Yang}, {Xue}, {Luo}, {Brandt}, {Alexander},
  {Bauer}, {Cui}, {Kong}, {Lehmer}, {Wang}, {Wu}, {Yuan}, {Yuan}, \&
  {Zhou}}]{yang14}
{Yang}, G., {Xue}, Y.~Q., {Luo}, B., {et~al.} 2014, \apjs, 215, 27,
  \dodoi{10.1088/0067-0049/215/2/27}

\bibitem[{{Younger} {et~al.}(2007){Younger}, {Huang}, {Fazio}, {Cox}, {Lai},
  {Hopkins}, {Hernquist}, {Papovich}, {Simard}, {Lin}, {Cheng}, {Yan}, {Kere{\v
  s}}, \& {Shapley}}]{younger07}
{Younger}, J.~D., {Huang}, J.-S., {Fazio}, G.~G., {et~al.} 2007, \apj, 671,
  1241, \dodoi{10.1086/522773}

\end{thebibliography}

\end{document}